\documentclass[dvipsnames,aps,pre,nobibnotes,twocolumn,superscriptaddress,nolongbibliography]{revtex4-1}

\usepackage[T1]{fontenc}
\usepackage[english]{babel}
\usepackage[utf8]{inputenc}

\usepackage[unicode=true,pdfusetitle,bookmarks=true,bookmarksnumbered=false,bookmarksopen=false,
 breaklinks=false,pdfborder={0 0 0},pdfborderstyle={},backref=false,colorlinks=true]
 {hyperref}
 \hypersetup{
 citecolor=blue,
 urlcolor=blue,
 linkcolor=blue}

\usepackage{array}
\usepackage{units}
\usepackage{amsmath}
\usepackage{cancel}
\usepackage{mathrsfs}
\usepackage{amssymb}
\usepackage{graphicx}
\usepackage{stackrel}
\usepackage{babel}
\usepackage{xparse}
\usepackage{soul}
\usepackage[final]{pdfpages}
\usepackage{comment}
\usepackage{tcolorbox}

\makeatletter
\AtBeginDocument{\let\LS@rot\@undefined}
\makeatother

\begin{document}
\title{Strange Attractors in Complex Networks}

\author{Pablo Villegas}
\email{pablo.villegas@cref.it}
\affiliation{`Enrico Fermi' Research Center (CREF), Via Panisperna 89A, 00184 - Rome, Italy}
\affiliation{Instituto Carlos I de F\'isica Te\'orica y Computacional, Univ. de Granada, E-18071, Granada, Spain.}

\begin{abstract}
Disorder and noise in physical systems often disrupt spatial and temporal regularity, yet chaotic systems reveal how order can emerge from unpredictable behavior. Complex networks, spatial analogs of chaos, exhibit disordered, non-Euclidean architectures with hidden symmetries, hinting at spontaneous order. Finding low-dimensional embeddings that reveal network patterns and link them to dimensionality that governs universal behavior remains a fundamental open challenge, as it needs to bridge the gap between microscopic disorder and macroscopic regularities. Here, the minimal space revealing key network properties is introduced, showing that non-integer dimensions produce chaotic-like attractors.
\end{abstract}

\maketitle  

Dimensionality and symmetries play a fundamental role that underlie universal behavior in physical systems. In statistical mechanics, it determines the corrections to mean-field critical exponents for the different dynamical universality classes \cite{MaBook,HenkelBook}, defines the relevance of quenched disorder \cite{ImryMa} or the potential spontaneous broken of continuous symmetries \cite{Mermin-Wagner}. In string theory, extra-dimensions are needed for theoretically unifying fundamental forces and particles \cite{Zwiebach2004}, while in nonlinear waves, are the lower dimensions that support interesting effects, such as integrability \cite{Falsi2024}.

Specifically, in the context of path integrals and field theory in statistical physics, the action determines the probabilities of different dynamical configurations or paths that a system can explore. The action is then a functional encapsulating the entire evolution of a system, analyzed by renormalization group methods at criticality, for Euclidean spaces \cite{OdorRMP,HenkelBook}. Hence, in fundamental universality classes, interactions between units satisfy minimal requirements of locality and homogeneity, leading the Laplacian operator to define their Gaussian approximation above the upper critical dimensions \cite{Hinrichsen2000, Harris1974, JSTAT}.%, and low spatial dimensionalities introduce perturbative corrections due to non-Gaussian effects \cite{AmitBook}. %which led to the original formulation of the LRG \cite{LRG, JSTAT}.

Complex networks comprise a fundamental class of disordered systems that often arise in the absence of Euclidean embedding spaces. For example, some structures have been shown to exhibit fractal dimensionalities greater than that of their growing space \cite{Daqing2011}. Identifying thus network patterns involves characterizing their local or global non-integer dimensionality \cite{Scale-invariance}, as this is key to uncovering underlying topological recurrences \cite{Peach2022, Samir-FSS, Debian}. This represents a crucial problem, as networks are the structural backbone of many real processes. Detecting non-integer dimensionalities in networks typically grounds on the Euclidean strategy of analyzing how the number of neighbors scales at varying distances \cite{Goh,Song}, which can lead to multiple problems due to small-world effects \cite{Cohen2003, WS, Rozenfeld2007,Rozenfeld2010}. Alternatively, random-walk-based dynamical approaches have used their temporal trajectories to effectively estimate their correlation dimension \cite{Lacasa}.

A precise definition of scale-invariant networks has been recently introduced using the Laplacian Renormalization Group (LRG) \cite{LRG}, linking the concept of non-integer dimension of a network to a constant entropy-loss rate across scales \cite{Scale-invariance}. This has enabled the differentiation between scale-free and scale-invariant networks while proposing structural universality classes. These classes comprise networks with a well-defined \emph{spectral dimension}, \(d_S\), a global property of the graph, related, e.g., to the infrared singularity of the Gaussian process \cite{Cassi1996} and that has been shown to provide a robust generalization of the standard concept of Euclidean dimension for heterogeneous systems \cite{Cassi1992,Cassi1996}. However, a multiscale framework such as the LRG needs to use the full matrix exponential or diagonalization to unravel intrinsic mesoscopic structures \cite{Modularity}, thus currently making the problem unfeasible for extensive systems. 

A basic open question is then to find embedding spaces where disordered systems can be arbitrarily represented to easily detect hidden symmetries and mesoscopic properties \cite{Gu2021, Boguna2021}. The hyperbolic embedding of complex networks has been proposed as an elegant and rigorous mathematical solution to this problem \cite{Hyper1,Hyper2,Boguna2021}, allowing to explain common structural properties, their navigability \cite{Boguna2009, Gulyas2015}, or efficient embeddings to be used in downstream tasks \cite{Almagro2022}. However, the definition of a low-dimensional space of embedding related to the physical dimension, the spectral one \cite{Cassi1992,Cassi1996}, still represents a fundamental unresolved issue.

In this letter, the minimum low-dimensional space where a network can be embedded to reveal underlying ordered patterns is presented. This naturally arises in reciprocal space, defined by the discrete network Laplacian. Moreover, the number of relevant eigenvectors presents a direct connection with the spectral dimension of the network. Such a representation sheds light on the emergence of hereto unknown strange attractors, the spatial analog of chaotic attractors, for scale-invariant and real networks. This offers a general solution for detecting hidden symmetries and developing in-depth analysis of previously intractable large networks.
\newpage

\paragraph*{\textbf{Low-dimensional embedding of heterogeneous systems.}}\hspace{-3mm}~ The perturbative renormalization group \cite{KardarBook} makes it natural to speculate that the discrete counterpart of $\nabla^2$, the network Laplacian, $\hat L= \hat D- \hat A$ (where $\hat D$ is the diagonal matrix and $\hat A$ the adjacency matrix of the weighted undirected network), is the operator containing all the relevant information of homogeneous and heterogeneous architectures. For example, the set of Laplacian eigenvalues contains the information on its intrinsic dimension under finite-size scale transformations \cite{Scale-invariance}.

Since $\hat L$ is a Hermitian, symmetric, and semi-positive defined operator, it can be diagonalized in a set of orthogonal vectors, being $|\lambda_j\rangle$ the ket associated with the $\lambda_j$ eigenvalue. Hence, the complete set, $j=1,...,N$, forms the orthogonal eigenbasis of $\hat L$ where each node can be naturally projected \cite{Belkin}. The ordered set of eigenvalues represents the network modes analogous to frequencies in Fourier space \cite{LRG}. In this framework, the eigenvalues correspond to different scales of oscillation, where lower eigenvalues relate to large-scale or low-frequency modes, and higher eigenvalues correspond to short-scale, high-frequency modes. In principle, the $k-$space defines an N-dimensional space for embedding any undirected and weighted network. Note that, for any connected network, $\lambda_0$ is the null eigenvalue and $|\lambda_0\rangle$ the uniform vector,  which is uninformative and can be safely discarded.

\begin{figure}[hbtp]
    \centering
    \includegraphics[width=1.0\columnwidth]{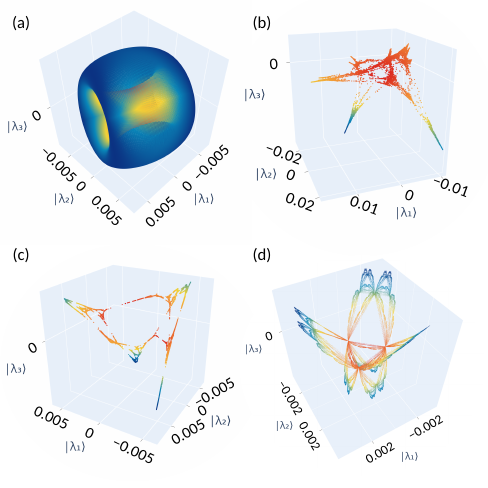}
    \caption{\textbf{3D embeddings.} Low-dimensional representation using the three smallest low-frequency normalized network modes, $|\lambda_1\rangle$, $|\lambda_2\rangle$, and $|\lambda_3\rangle$ as the coordinate axes for: \textbf{(a)} A 2D lattice of size $L=256$. \textbf{(b)} A KH network of size $N=2\cdot10^4$ with $m=3$ and $p=1$. \textbf{(c)} A HMN network with $s=14,\alpha=3,$, and $M_0=3$ consisting on $N=49152$ nodes. \textbf{(d)} A DGM network of size $N=256722$. }
    \label{3DEmbedd}
\end{figure}

Fig.\ref{3DEmbedd} shows the three-dimensional embedding in the space defined by the three first low-frequency network modes (see Supplemental Material (SM) \cite{SM} for different examples of 2D planar projections) for a bi-dimensional squared lattice with periodic boundary conditions (see Fig.\ref{3DEmbedd}(a)), and different networks with well-defined scale-invariant properties and non-integer dimensions recently described in \cite{Scale-invariance} (see SM \cite{SM} for further details on how these networks are built), as the one proposed by Kim and Holme (KH) \cite{KimHolme} (see Fig.\ref{3DEmbedd}(b)), a Hierarchical Modular Network \cite{moretti2013griffiths,SciRep} (HMN, see Fig.\ref{3DEmbedd}(c)), and the Dorogovtsev-Goltsev-Mendes one \cite{DGM} (DGM, see Fig.\ref{3DEmbedd}(d)). The first consequence of a well-defined non-integer dimensionality is that this low-dimensional projection exhibits a strange, chaotic-like attractor for the system. Instead, regular structures maintain their intrinsic properties, such as occurs in reciprocal lattices \cite{Kittel2005}, and being tantamount to far-field optical imaging \cite{Xin2024}. 

\paragraph*{\textbf{Correlation dimension of complex networks.}} The correlation integral proposed by Grassberger and Procaccia \cite{CorrDim,ScaleFree} allows to address the problem of analyzing the minimal embedding space. Consider the correlation sum around the $\mathbf{x}_i$ point, namely,
\begin{equation}
    C(\ell)=\frac{1}{N-1} \sum_{i=1}^{N_{c}(\ell)}\frac{n_\ell(\mathbf{x}_i)}{N_{c}(\ell)},
\end{equation}
where $N_{c}(\ell)$ are the number of valid centers up to scale $\ell$ in reciprocal space, being each point expressed as a linear combination of the considered eigenvectors, $|i\rangle = \sum_{k=1}^{d} |\lambda_k\rangle \langle \lambda_k | i \rangle$ with \(\mathbf{x}_i = \left( \langle \lambda_1 | i \rangle, \langle \lambda_2 | i \rangle, \ldots, \langle \lambda_d | i \rangle \right)\) the projection of the $i^{th}$ node onto the d-dimensional eigenbasis formed by the first d eigenvectors of the Laplacian. This defines the dimension $d$ of the embedding space. In particular, if $C(\ell)$ scales as a power law, $C(\ell) \propto \ell^D$, $D$ defines the network correlation dimension. 

\begin{figure}[hbtp]
    \centering
    \includegraphics[width=1.0\columnwidth]{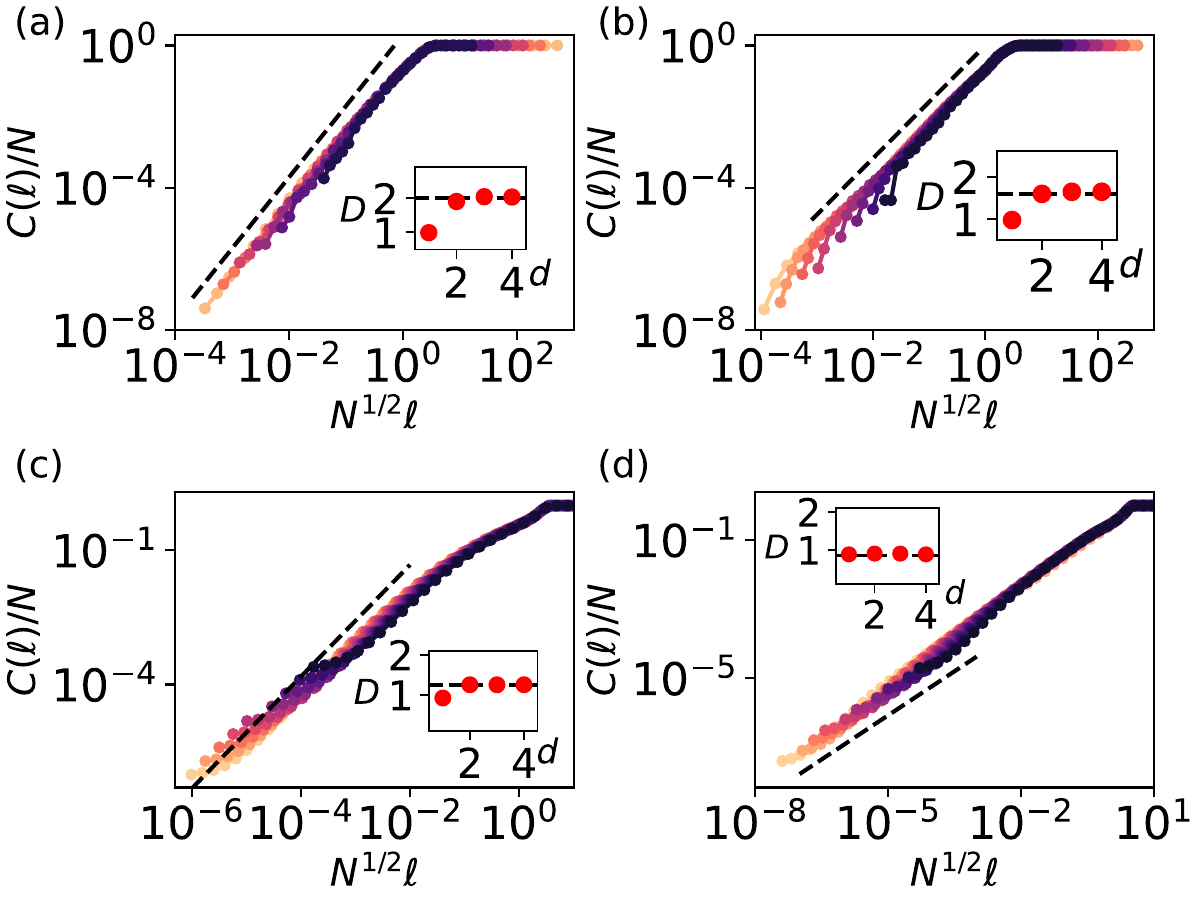}
    \caption{Correlation integral, $C(\ell)$ versus normalized distance, $\sqrt{N}\ell$, for different sizes. Inset: Estimated correlation dimension $(D)$ versus embedding dimension $(d)$. \textbf{(a)} 2D lattice of size $N=L^2$, with $L=16,32,64,128,256$ and $512$. \textbf{(b)} DGM networks of size $N=\nicefrac{3+3^s}{2}$ with $s=6,7,8,9,10,11,12$ and $13$. \textbf{(b)} KH networks with $m=2$ and $p=1$ of size $N=2^s$ and $s=10,11,12,13,14,15,16,17$ and $18$. \textbf{(c)} HMN networks with $\alpha=3$, and $M_0=3$, of size $N=M_02^s$ with $s=9,10,11,12,13,14,15,16,17$ and $18$. HMN and KH networks have been averaged over $10^2-10^3$ independent realizations. Black dashed lines represent the scaling as predicted by $d_E$ for 2D lattices or $d_S/2$ for the others.}
    \label{CorrInt}
\end{figure}

The careful analysis of $C(\ell)$ thus allows to analyze how the correlation dimension changes with the dimension of the embedding space. In particular, the results for a 2D lattice are reported in Fig.\ref{CorrInt}(a). Not surprisingly, the correlation dimension is well-defined from $d=2$ onward. 
Since this accounts for the d-dimensional projection of a D-dimensional structure, the attractor of a network can be measured only for $D<d$ \cite{Falconer2014,Rams2014,Falsi2021}, limiting the available 'degrees of freedom' and leading to define the minimal space to embed the network: the immediate one after the network correlation dimension, $d_{\min}=\lceil D \rceil$.

Fig.\ref{CorrInt}(b) shows the correlation integral for the case of a DGM network. This network presents a power-law degree distribution, high clustering coefficient, and small-world properties with discrete scale invariance \cite{DGM}. In particular, the spectral dimension is $d_S=\frac{2\log(3)}{\log(2)}\approx3.17$, which corresponds to half of the measured correlation dimension $D$. Consequently, the network can be safely embedded from dimension $d=2$ onward. Fig.\ref{CorrInt}(c) shows the case of a KH network, a \emph{stochastic} scale-invariant network with $d_S=2.57(1)$ \cite{Scale-invariance}. Even if the correlation integral fits well for half of the spectral dimension at small values of $\ell$, it presents small deviations at large $\ell-$values (see below). This is equivalent to the usually detected 'knee', which appears in chaotic systems if the signal from a deterministic chaotic system is perturbed by adding random noise of smaller amplitude \cite{Eckmann1985, Argyrys1998}.
Finally, for HMN networks, Fig.\ref{CorrInt}(d) reports the scaling compatible with the expected spectral dimension, $d_S=1.70(1)$ of the network. See also SM for further examples with random trees and further analysis with different values of the embedding dimension.

The direct observation of the correlation dimension illustrates the difference between the attractors of systems that naturally live in a well-defined Euclidean dimension (e.g., lattices) and those with a non-integer dimension as scale-invariant networks. In particular, the correlation dimension matches with the Euclidean one ($D=d_E$) when representing a regular system or a lattice, while it strictly coincides with half of the spectral dimension for heterogeneous systems ($D=d_S/2$). In fact, the scaling of the Fiedler eigenvalue (i.e., the infrared behavior of the network) as a function of the system size, $\lambda_F\sim N^{-2/d_S}$, represents just the scalar projection of the actual attractors. The KH and HMN networks analysis reveals that such a scalar projection can still lose some information regarding the strange attractor characterizing complex networks (for instance, the stochastic effects at large $\ell$, or multiple scaling regions, see below). 

\begin{figure}
    \centering
    \includegraphics[width=1.0\columnwidth]{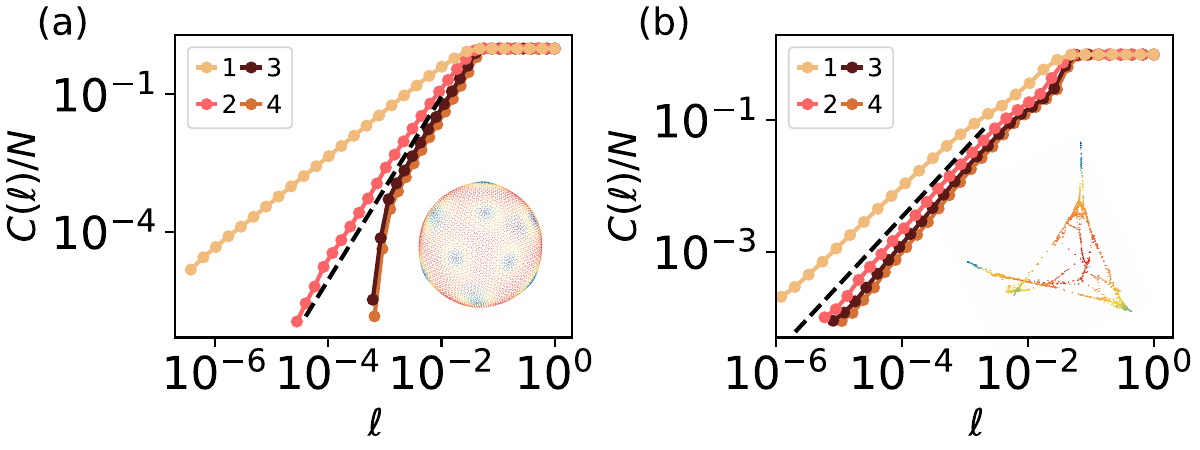}
    \caption{Correlation integral $C(\ell)$ versus distance $\ell$, for different embedding spaces (see legend). Black dashed lines are guides to the eye to the fitted correlation dimension, $D$. Insets show the 3D representation of the networks using the three smallest low-frequency normalized network modes. \textbf{(a)} Fullerene network $C_{6000}$. For $d\geq2$, the structures show a well-defined scaling region with $D=d_E=2$. \textbf{(b)} US power grid network. The network exhibits a well-defined correlation dimension $D=d_S/2=1.02(4)$ for $d\geq2$.}
    \label{Spatial}
\end{figure}

The analysis is now extended to different real-world networks. First, the fullerene graphs are considered, where nodes and edges correspond to atoms and bonds of a molecule, respectively \cite{Aref2019}. Fig.\ref{Spatial}(a) confirms the 2D nature of these structures, with a correlation dimension $D=d_E=2$. Notably, the embedding successfully recovers the 3D structure of the fullerene by considering only its adjacency matrix, acting as a test bed for evaluating the effectiveness in capturing topological and spatial features.  Instead, Fig.\ref{Spatial}(b) reports the correlation integral for the US power grid originally used to analyze small-world phenomena \cite{WS}. Here, vertices represent generators, transformers, and substations, and edges represent high-voltage transmission lines between them. Indeed, as emerges from the analysis, this network presents a dimension $d_S=2D=2.05$, which reflects the 2D space where the network has grown. Moreover, different results regarding the road networks of the 50 US states and the European road network demonstrate (i) the ongoing applicability of the method to extensive networks up to more than $10^6$ nodes and (ii) the presence of a dimension in road networks from $d_S=1$ to $d_S=2$, which is fully compatible to previous results \cite{Lu2004}. One important issue that has not been discussed before is the ability to capture the loss of the spatial inherent nature of the system,  i.e., $D=d_E$, instead of $D=d_S/2$, for some US states depending on the intrinsic structural heterogeneity (see SM \cite{SM}). These examples were not hand-picked; they were chosen because of their inherent interest, giving real structures for testing the previous hypothesis.

\begin{figure}
    \centering
    \includegraphics[width=1.0\columnwidth]{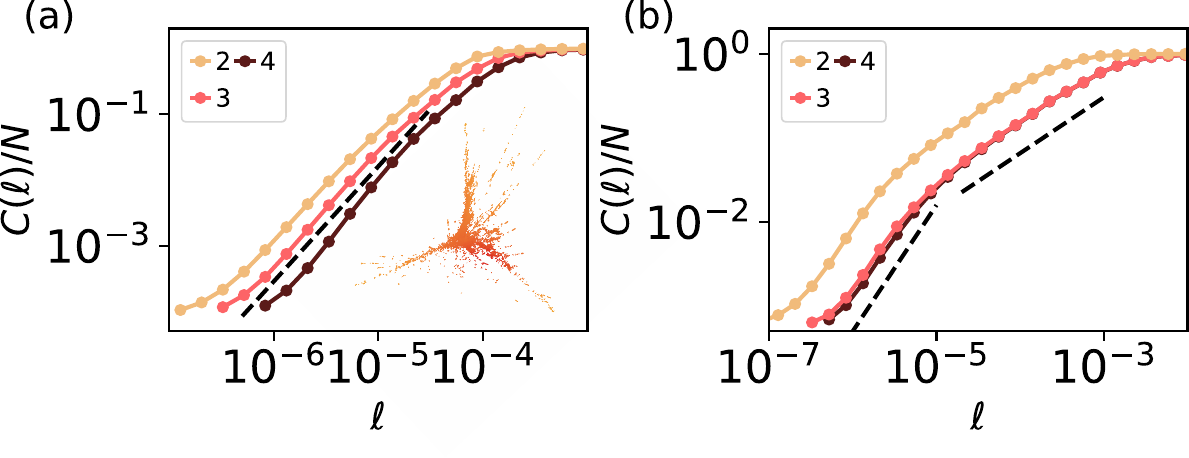}
    \caption{Correlation integral $C(\ell)$ versus distance $\ell$, for different embedding spaces (see legend). Black dashed lines are guides to the eye to the fitted correlation dimension, $D$. Insets show the 3D representation of the networks using the three smallest low-frequency normalized network modes. \textbf{(a)} Nematode parasites-mammals network. For $d\geq2$, the structures show a well-defined scaling region with $d_S=3.5(1)$. \textbf{(b)} C. Elegans PPI network. The network exhibits two scaling regions, one with $d_S\approx1.33$ and another one with $d_S\approx3.0$.}
    \label{LastNets}
\end{figure}

Fig.\ref{LastNets} reports the analysis of the bipartite network of interactions between nematodes and their host mammal species \cite{Dallas2016,Dallas2018} and the \emph{Caenorhabditis elegans} protein-protein interactome network \cite{Simonis2009}, two cases of high-complexity. The first one reveals that this bipartite network has a finite dimension $d_S\approx3.5(1)$ (see Fig.\ref{LastNets}(a)). This can reflect the existence of well-defined scaling relationships between species richness and the geographical area sampled in these networks \cite{Dallas2018}. Otherwise, the C. Elegans PPI network allows the detection of two scaling regions, one with $d_S\approx3$ and one with $d_S\approx1.33$. This is particularly relevant because the scales are being scrutinized in reciprocal space \cite{LRG}: at the local scale (small $k-values$), the network exhibits a tree-like structure, but at large scales (large $k-values$), the network presents a three-dimensional structure. In fact, this recalls the two types of molecular assemblies required for embryogenesis: multi-protein complexes like the ribosome or anaphase-promoting complex and highly connected subnetworks of proteins acting in distinct but functionally interdependent processes \cite{Simonis2009, Gunsalus2005}. Moreover, in the case of the previously analyzed KH networks, it naturally explains why the 'knee' appears; it corresponds to stochastic fluctuations that affect local scales.

Finally, the analysis of stochastic block models (see SM \cite{SM}) reveals how their embedding allows the identification of mesoscopic structures as different densely populated areas in reciprocal space. This extends previous approaches \cite{Donetti} to $d$ dimensions where the reduced overlap probability facilitates their detection.

\paragraph*{\textbf{Outlook.}} 
Order is the foundation of communication between elements at any level of organization, whether that refers to a population of neurons, bees, or humans \cite{Crut2012}. At the same time, disorder seems to be an intrinsic and unavoidable feature of real systems \cite{Strogatz2001}, where interactions between units form heterogeneous and intricate patterns that are not apparently related to the physical space where they dynamically operate \cite{Doro2008}. 

Here, the low-dimensional projection of general heterogeneous networks in reciprocal space is analyzed. While network connections are not directly related to natural manifolds or coordinate systems, the reciprocal space defined by the Laplacian eigenvectors illustrates the existence of symmetries and ordered patterns in such structures. In particular, this provides the network Fourier basis for regular and heterogeneous networks. It naturally represents a coordinate system where the network can be embedded in a space that will depend on its intrinsic dimensionality. This dimensionality can be captured through standard analysis as the correlation integral proposed by Grassberger and Procaccia to analyze the dimensionality of chaotic and strange attractors \cite{CorrDim}. Moreover, the embedding of regular and heterogeneous networks reveals that, while regular networks maintain their structure, scale-invariant networks exhibit a strange chaotic-like attractor with dimension half of the spectral dimension of the network. The emergence of strange attractors encoding hidden symmetries seems not to be merely a curiosity of spatial networks \cite{Barth2011} nor an artifact of idealized models: it is probably generic for many large, sparse networks found in nature. The results presented here also solve a fundamental open debate about the natural space dimension one should choose for embedding a heterogeneous system \cite{Boguna2021}.

An important consequence, due to the difficulty of finding scale-invariant networks with $d_S>4$ \cite{Scale-invariance}, is the reliability of 3D embeddings in providing a robust framework for the decomposition of most networks. Such 3D embeddings also enable an Euclidean-like representation of these highly disordered structures, facilitating visual inspection of their properties. Also, even if the LRG has recently shed light on true structural scale invariance \cite{Scale-invariance}, the presented results make detecting particular network resolution scales and regularities possible in real cases where fluctuations across such intertwined scales make it extremely difficult to rigorously maintain scale invariance or it partially appears in some network regions. Still, this highlights an important question that remains open: the current challenge of theoretically proposing high-dimensional scale-invariant networks with finite spectral dimension $d_S\geq4$. 

Finally, while the LRG framework provides the complete solution for scrutinizing the full range of scales of the network by considering the entire set of eigenvalues \cite{LRG,Modularity}, the low-dimensional Laplacian embedding permits a robust and computationally efficient version of the community detection algorithm recently proposed in \cite{Modularity}. This is also a direct extension of previous proposed spectral methods \cite{Donetti}, which do not require any null model to detect modules as they only consider dense areas of the embedding space to group nodes.

The findings presented here open a route for studying multiple biological and socio-technological networks in 2D and 3D embedding spaces. However, the formal connection with actual hyperbolic embeddings remains a crucial question to be solved for understanding the geometry and fundamental symmetries of networks \cite{Hyper2,Boguna2021}.
\vspace{-0.85cm}
\begin{acknowledgments}
\vspace{-0.25cm}
I thank L. Falsi and A. Gabrielli for very useful discussions and comments. I acknowledge the Spanish Ministry of Research and Innovation and Agencia Estatal de Investigación (AEI), MICIN/AEI/10.13039/501100011033, for financial support through Project PID2023-149174NB-I00, funded also by European Regional Development Funds, and  Ref. PID2020-113681GB-I00.
\newpage

\end{acknowledgments}

\section*{Data availability}
All real networks analyzed in this paper are freely available at the Netzschleuder database \cite{Peixoto2024}. The Python programs, in the form of Jupyter notebooks, related to this article are available in the GitHub repository \cite{GH}.
\def\url#1{}
%\bibliography{Attractors}

\begin{thebibliography}{62}%
\makeatletter
\providecommand \@ifxundefined [1]{%
 \@ifx{#1\undefined}
}%
\providecommand \@ifnum [1]{%
 \ifnum #1\expandafter \@firstoftwo
 \else \expandafter \@secondoftwo
 \fi
}%
\providecommand \@ifx [1]{%
 \ifx #1\expandafter \@firstoftwo
 \else \expandafter \@secondoftwo
 \fi
}%
\providecommand \natexlab [1]{#1}%
\providecommand \enquote  [1]{``#1''}%
\providecommand \bibnamefont  [1]{#1}%
\providecommand \bibfnamefont [1]{#1}%
\providecommand \citenamefont [1]{#1}%
\providecommand \href@noop [0]{\@secondoftwo}%
\providecommand \href [0]{\begingroup \@sanitize@url \@href}%
\providecommand \@href[1]{\@@startlink{#1}\@@href}%
\providecommand \@@href[1]{\endgroup#1\@@endlink}%
\providecommand \@sanitize@url [0]{\catcode `\\12\catcode `\$12\catcode
  `\&12\catcode `\#12\catcode `\^12\catcode `\_12\catcode `\%12\relax}%
\providecommand \@@startlink[1]{}%
\providecommand \@@endlink[0]{}%
\providecommand \url  [0]{\begingroup\@sanitize@url \@url }%
\providecommand \@url [1]{\endgroup\@href {#1}{\urlprefix }}%
\providecommand \urlprefix  [0]{URL }%
\providecommand \Eprint [0]{\href }%
\providecommand \doibase [0]{http://dx.doi.org/}%
\providecommand \selectlanguage [0]{\@gobble}%
\providecommand \bibinfo  [0]{\@secondoftwo}%
\providecommand \bibfield  [0]{\@secondoftwo}%
\providecommand \translation [1]{[#1]}%
\providecommand \BibitemOpen [0]{}%
\providecommand \bibitemStop [0]{}%
\providecommand \bibitemNoStop [0]{.\EOS\space}%
\providecommand \EOS [0]{\spacefactor3000\relax}%
\providecommand \BibitemShut  [1]{\csname bibitem#1\endcsname}%
\let\auto@bib@innerbib\@empty
%</preamble>
\bibitem [{\citenamefont {Ma}(2018)}]{MaBook}%
  \BibitemOpen
  \bibfield  {author} {\bibinfo {author} {\bibfnamefont {S.-K.}\ \bibnamefont
  {Ma}},\ }\href@noop {} {\emph {\bibinfo {title} {Modern Theory of Critical
  Phenomena}}}\ (\bibinfo  {publisher} {Routledge},\ \bibinfo {address} {New
  York},\ \bibinfo {year} {2018})\BibitemShut {NoStop}%
\bibitem [{\citenamefont {Henkel}\ \emph {et~al.}(2008)\citenamefont {Henkel},
  \citenamefont {Hinrichsen},\ and\ \citenamefont {Lübeck}}]{HenkelBook}%
  \BibitemOpen
  \bibfield  {author} {\bibinfo {author} {\bibfnamefont {M.}~\bibnamefont
  {Henkel}}, \bibinfo {author} {\bibfnamefont {H.}~\bibnamefont {Hinrichsen}},
  \ and\ \bibinfo {author} {\bibfnamefont {S.}~\bibnamefont {Lübeck}},\
  }\href@noop {} {\emph {\bibinfo {title} {Non-Equilibrium Phase Transitions:
  Volume 1: Absorbing Phase Transitions}}}\ (\bibinfo  {publisher} {Springer},\
  \bibinfo {address} {Dordrecht},\ \bibinfo {year} {2008})\BibitemShut
  {NoStop}%
\bibitem [{\citenamefont {Imry}\ and\ \citenamefont {Ma}(1975)}]{ImryMa}%
  \BibitemOpen
  \bibfield  {author} {\bibinfo {author} {\bibfnamefont {Y.}~\bibnamefont
  {Imry}}\ and\ \bibinfo {author} {\bibfnamefont {S.-k.}\ \bibnamefont {Ma}},\
  }\href {\doibase 10.1103/PhysRevLett.35.1399} {\bibfield  {journal} {\bibinfo
   {journal} {Phys. Rev. Lett.}\ }\textbf {\bibinfo {volume} {35}},\ \bibinfo
  {pages} {1399} (\bibinfo {year} {1975})}\BibitemShut {NoStop}%
\bibitem [{\citenamefont {Mermin}\ and\ \citenamefont
  {Wagner}(1966)}]{Mermin-Wagner}%
  \BibitemOpen
  \bibfield  {author} {\bibinfo {author} {\bibfnamefont {N.~D.}\ \bibnamefont
  {Mermin}}\ and\ \bibinfo {author} {\bibfnamefont {H.}~\bibnamefont
  {Wagner}},\ }\href {\doibase 10.1103/PhysRevLett.17.1133} {\bibfield
  {journal} {\bibinfo  {journal} {Phys. Rev. Lett.}\ }\textbf {\bibinfo
  {volume} {17}},\ \bibinfo {pages} {1133} (\bibinfo {year}
  {1966})}\BibitemShut {NoStop}%
\bibitem [{\citenamefont {Zwiebach}(2004)}]{Zwiebach2004}%
  \BibitemOpen
  \bibfield  {author} {\bibinfo {author} {\bibfnamefont {B.}~\bibnamefont
  {Zwiebach}},\ }\href@noop {} {\emph {\bibinfo {title} {A First Course in
  String Theory}}}\ (\bibinfo  {publisher} {Cambridge University Press},\
  \bibinfo {address} {Cambridge},\ \bibinfo {year} {2004})\BibitemShut
  {NoStop}%
\bibitem [{\citenamefont {Falsi}\ \emph {et~al.}(2024)\citenamefont {Falsi},
  \citenamefont {Villois}, \citenamefont {Coppini}, \citenamefont {Agranat},
  \citenamefont {DelRe},\ and\ \citenamefont {Trillo}}]{Falsi2024}%
  \BibitemOpen
  \bibfield  {author} {\bibinfo {author} {\bibfnamefont {L.}~\bibnamefont
  {Falsi}}, \bibinfo {author} {\bibfnamefont {A.}~\bibnamefont {Villois}},
  \bibinfo {author} {\bibfnamefont {F.}~\bibnamefont {Coppini}}, \bibinfo
  {author} {\bibfnamefont {A.~J.}\ \bibnamefont {Agranat}}, \bibinfo {author}
  {\bibfnamefont {E.}~\bibnamefont {DelRe}}, \ and\ \bibinfo {author}
  {\bibfnamefont {S.}~\bibnamefont {Trillo}},\ }\href {\doibase
  10.1103/PhysRevLett.133.183804} {\bibfield  {journal} {\bibinfo  {journal}
  {Phys. Rev. Lett.}\ }\textbf {\bibinfo {volume} {133}},\ \bibinfo {pages}
  {183804} (\bibinfo {year} {2024})}\BibitemShut {NoStop}%
\bibitem [{\citenamefont {\'Odor}(2004)}]{OdorRMP}%
  \BibitemOpen
  \bibfield  {author} {\bibinfo {author} {\bibfnamefont {G.}~\bibnamefont
  {\'Odor}},\ }\href {\doibase 10.1103/RevModPhys.76.663} {\bibfield  {journal}
  {\bibinfo  {journal} {Rev. Mod. Phys.}\ }\textbf {\bibinfo {volume} {76}},\
  \bibinfo {pages} {663} (\bibinfo {year} {2004})}\BibitemShut {NoStop}%
\bibitem [{\citenamefont {Hinrichsen}(2000)}]{Hinrichsen2000}%
  \BibitemOpen
  \bibfield  {author} {\bibinfo {author} {\bibfnamefont {H.}~\bibnamefont
  {Hinrichsen}},\ }\href {\doibase 10.1080/00018730050198152} {\bibfield
  {journal} {\bibinfo  {journal} {Adv. Phys.}\ }\textbf {\bibinfo {volume}
  {49}},\ \bibinfo {pages} {815} (\bibinfo {year} {2000})}\BibitemShut
  {NoStop}%
\bibitem [{\citenamefont {Harris}(1974)}]{Harris1974}%
  \BibitemOpen
  \bibfield  {author} {\bibinfo {author} {\bibfnamefont {A.~B.}\ \bibnamefont
  {Harris}},\ }\href {\doibase 10.1088/0022-3719/7/9/009} {\bibfield  {journal}
  {\bibinfo  {journal} {J. Phys. C: Solid State Phys.}\ }\textbf {\bibinfo
  {volume} {7}},\ \bibinfo {pages} {1671} (\bibinfo {year} {1974})}\BibitemShut
  {NoStop}%
\bibitem [{\citenamefont {Caldarelli}\ \emph {et~al.}(2024)\citenamefont
  {Caldarelli}, \citenamefont {Gabrielli}, \citenamefont {Gili},\ and\
  \citenamefont {Villegas}}]{JSTAT}%
  \BibitemOpen
  \bibfield  {author} {\bibinfo {author} {\bibfnamefont {G.}~\bibnamefont
  {Caldarelli}}, \bibinfo {author} {\bibfnamefont {A.}~\bibnamefont
  {Gabrielli}}, \bibinfo {author} {\bibfnamefont {T.}~\bibnamefont {Gili}}, \
  and\ \bibinfo {author} {\bibfnamefont {P.}~\bibnamefont {Villegas}},\ }\href
  {\doibase 10.1088/1742-5468/ad57b1} {\bibfield  {journal} {\bibinfo
  {journal} {J. Stat. Mech.: Theory Exp.}\ }\textbf {\bibinfo {volume}
  {2024}},\ \bibinfo {pages} {084002} (\bibinfo {year} {2024})}\BibitemShut
  {NoStop}%
\bibitem [{\citenamefont {Daqing}\ \emph {et~al.}(2011)\citenamefont {Daqing},
  \citenamefont {Kosmidis}, \citenamefont {Bunde},\ and\ \citenamefont
  {Havlin}}]{Daqing2011}%
  \BibitemOpen
  \bibfield  {author} {\bibinfo {author} {\bibfnamefont {L.}~\bibnamefont
  {Daqing}}, \bibinfo {author} {\bibfnamefont {K.}~\bibnamefont {Kosmidis}},
  \bibinfo {author} {\bibfnamefont {A.}~\bibnamefont {Bunde}}, \ and\ \bibinfo
  {author} {\bibfnamefont {S.}~\bibnamefont {Havlin}},\ }\href {\doibase
  10.1038/nphys1932} {\bibfield  {journal} {\bibinfo  {journal} {Nat. Phys.}\
  }\textbf {\bibinfo {volume} {7}},\ \bibinfo {pages} {481} (\bibinfo {year}
  {2011})}\BibitemShut {NoStop}%
\bibitem [{\citenamefont {Poggialini}\ \emph {et~al.}(2025)\citenamefont
  {Poggialini}, \citenamefont {Villegas}, \citenamefont {Mu\~noz},\ and\
  \citenamefont {Gabrielli}}]{Scale-invariance}%
  \BibitemOpen
  \bibfield  {author} {\bibinfo {author} {\bibfnamefont {A.}~\bibnamefont
  {Poggialini}}, \bibinfo {author} {\bibfnamefont {P.}~\bibnamefont
  {Villegas}}, \bibinfo {author} {\bibfnamefont {M.~A.}\ \bibnamefont
  {Mu\~noz}}, \ and\ \bibinfo {author} {\bibfnamefont {A.}~\bibnamefont
  {Gabrielli}},\ }\href {\doibase 10.1103/PhysRevLett.134.057401} {\bibfield
  {journal} {\bibinfo  {journal} {Phys. Rev. Lett.}\ }\textbf {\bibinfo
  {volume} {134}},\ \bibinfo {pages} {057401} (\bibinfo {year}
  {2025})}\BibitemShut {NoStop}%
\bibitem [{\citenamefont {Peach}\ \emph {et~al.}(2022)\citenamefont {Peach},
  \citenamefont {Arnaudon},\ and\ \citenamefont {Barahona}}]{Peach2022}%
  \BibitemOpen
  \bibfield  {author} {\bibinfo {author} {\bibfnamefont {R.}~\bibnamefont
  {Peach}}, \bibinfo {author} {\bibfnamefont {A.}~\bibnamefont {Arnaudon}}, \
  and\ \bibinfo {author} {\bibfnamefont {M.}~\bibnamefont {Barahona}},\ }\href
  {\doibase 10.1038/s41467-022-30705-w} {\bibfield  {journal} {\bibinfo
  {journal} {Nat. Comm.}\ }\textbf {\bibinfo {volume} {13}},\ \bibinfo {pages}
  {3088} (\bibinfo {year} {2022})}\BibitemShut {NoStop}%
\bibitem [{\citenamefont {Serafino}\ \emph {et~al.}(2021)\citenamefont
  {Serafino}, \citenamefont {Cimini}, \citenamefont {Maritan}, \citenamefont
  {Rinaldo}, \citenamefont {Suweis}, \citenamefont {Banavar},\ and\
  \citenamefont {Caldarelli}}]{Samir-FSS}%
  \BibitemOpen
  \bibfield  {author} {\bibinfo {author} {\bibfnamefont {M.}~\bibnamefont
  {Serafino}}, \bibinfo {author} {\bibfnamefont {G.}~\bibnamefont {Cimini}},
  \bibinfo {author} {\bibfnamefont {A.}~\bibnamefont {Maritan}}, \bibinfo
  {author} {\bibfnamefont {A.}~\bibnamefont {Rinaldo}}, \bibinfo {author}
  {\bibfnamefont {S.}~\bibnamefont {Suweis}}, \bibinfo {author} {\bibfnamefont
  {J.~R.}\ \bibnamefont {Banavar}}, \ and\ \bibinfo {author} {\bibfnamefont
  {G.}~\bibnamefont {Caldarelli}},\ }\href {\doibase 10.1073/pnas.2013825118}
  {\bibfield  {journal} {\bibinfo  {journal} {Proc. Natl. Acad. Sci. U.S. A.}\
  }\textbf {\bibinfo {volume} {118}},\ \bibinfo {pages} {e2013825118} (\bibinfo
  {year} {2021})}\BibitemShut {NoStop}%
\bibitem [{\citenamefont {Villegas}\ \emph {et~al.}(2020)\citenamefont
  {Villegas}, \citenamefont {Mu{\~n}oz},\ and\ \citenamefont
  {Bonachela}}]{Debian}%
  \BibitemOpen
  \bibfield  {author} {\bibinfo {author} {\bibfnamefont {P.}~\bibnamefont
  {Villegas}}, \bibinfo {author} {\bibfnamefont {M.~A.}\ \bibnamefont
  {Mu{\~n}oz}}, \ and\ \bibinfo {author} {\bibfnamefont {J.~A.}\ \bibnamefont
  {Bonachela}},\ }\href {\doibase 10.1098/rsif.2019.0845} {\bibfield  {journal}
  {\bibinfo  {journal} {J. R. Soc. Interface}\ }\textbf {\bibinfo {volume}
  {17}},\ \bibinfo {pages} {20190845} (\bibinfo {year} {2020})}\BibitemShut
  {NoStop}%
\bibitem [{\citenamefont {Goh}\ \emph {et~al.}(2006)\citenamefont {Goh},
  \citenamefont {Salvi}, \citenamefont {Kahng},\ and\ \citenamefont
  {Kim}}]{Goh}%
  \BibitemOpen
  \bibfield  {author} {\bibinfo {author} {\bibfnamefont {K.-I.}\ \bibnamefont
  {Goh}}, \bibinfo {author} {\bibfnamefont {G.}~\bibnamefont {Salvi}}, \bibinfo
  {author} {\bibfnamefont {B.}~\bibnamefont {Kahng}}, \ and\ \bibinfo {author}
  {\bibfnamefont {D.}~\bibnamefont {Kim}},\ }\href {\doibase
  10.1103/PhysRevLett.96.018701} {\bibfield  {journal} {\bibinfo  {journal}
  {Phys. Rev. Lett.}\ }\textbf {\bibinfo {volume} {96}},\ \bibinfo {pages}
  {018701} (\bibinfo {year} {2006})}\BibitemShut {NoStop}%
\bibitem [{\citenamefont {Song}\ \emph {et~al.}(2005)\citenamefont {Song},
  \citenamefont {Havlin},\ and\ \citenamefont {Makse}}]{Song}%
  \BibitemOpen
  \bibfield  {author} {\bibinfo {author} {\bibfnamefont {C.}~\bibnamefont
  {Song}}, \bibinfo {author} {\bibfnamefont {S.}~\bibnamefont {Havlin}}, \ and\
  \bibinfo {author} {\bibfnamefont {H.~A.}\ \bibnamefont {Makse}},\ }\href
  {\doibase doi.org/10.1038/nature03248} {\bibfield  {journal} {\bibinfo
  {journal} {Nature}\ }\textbf {\bibinfo {volume} {433}},\ \bibinfo {pages}
  {392} (\bibinfo {year} {2005})}\BibitemShut {NoStop}%
\bibitem [{\citenamefont {Cohen}\ and\ \citenamefont
  {Havlin}(2003)}]{Cohen2003}%
  \BibitemOpen
  \bibfield  {author} {\bibinfo {author} {\bibfnamefont {R.}~\bibnamefont
  {Cohen}}\ and\ \bibinfo {author} {\bibfnamefont {S.}~\bibnamefont {Havlin}},\
  }\href {\doibase 10.1103/PhysRevLett.90.058701} {\bibfield  {journal}
  {\bibinfo  {journal} {Phys. Rev. Lett.}\ }\textbf {\bibinfo {volume} {90}},\
  \bibinfo {pages} {058701} (\bibinfo {year} {2003})}\BibitemShut {NoStop}%
\bibitem [{\citenamefont {Watts}\ and\ \citenamefont {Strogatz}(1998)}]{WS}%
  \BibitemOpen
  \bibfield  {author} {\bibinfo {author} {\bibfnamefont {D.~J.}\ \bibnamefont
  {Watts}}\ and\ \bibinfo {author} {\bibfnamefont {S.~H.}\ \bibnamefont
  {Strogatz}},\ }\href {\doibase 10.1038/30918} {\bibfield  {journal} {\bibinfo
   {journal} {Nature}\ }\textbf {\bibinfo {volume} {393}},\ \bibinfo {pages}
  {440} (\bibinfo {year} {1998})}\BibitemShut {NoStop}%
\bibitem [{\citenamefont {Rozenfeld}\ \emph {et~al.}(2007)\citenamefont
  {Rozenfeld}, \citenamefont {Havlin},\ and\ \citenamefont
  {Ben-Avraham}}]{Rozenfeld2007}%
  \BibitemOpen
  \bibfield  {author} {\bibinfo {author} {\bibfnamefont {H.~D.}\ \bibnamefont
  {Rozenfeld}}, \bibinfo {author} {\bibfnamefont {S.}~\bibnamefont {Havlin}}, \
  and\ \bibinfo {author} {\bibfnamefont {D.}~\bibnamefont {Ben-Avraham}},\
  }\href {\doibase 10.1088/1367-2630/9/6/175} {\bibfield  {journal} {\bibinfo
  {journal} {New J. Phys.}\ }\textbf {\bibinfo {volume} {9}},\ \bibinfo {pages}
  {175} (\bibinfo {year} {2007})}\BibitemShut {NoStop}%
\bibitem [{\citenamefont {Rozenfeld}\ \emph {et~al.}(2010)\citenamefont
  {Rozenfeld}, \citenamefont {Song},\ and\ \citenamefont
  {Makse}}]{Rozenfeld2010}%
  \BibitemOpen
  \bibfield  {author} {\bibinfo {author} {\bibfnamefont {H.~D.}\ \bibnamefont
  {Rozenfeld}}, \bibinfo {author} {\bibfnamefont {C.}~\bibnamefont {Song}}, \
  and\ \bibinfo {author} {\bibfnamefont {H.~A.}\ \bibnamefont {Makse}},\ }\href
  {\doibase 10.1103/PhysRevLett.104.025701} {\bibfield  {journal} {\bibinfo
  {journal} {Phys. Rev. Lett.}\ }\textbf {\bibinfo {volume} {104}},\ \bibinfo
  {pages} {025701} (\bibinfo {year} {2010})}\BibitemShut {NoStop}%
\bibitem [{\citenamefont {Lacasa}\ and\ \citenamefont
  {G\'omez-Garde\~nes}(2013)}]{Lacasa}%
  \BibitemOpen
  \bibfield  {author} {\bibinfo {author} {\bibfnamefont {L.}~\bibnamefont
  {Lacasa}}\ and\ \bibinfo {author} {\bibfnamefont {J.}~\bibnamefont
  {G\'omez-Garde\~nes}},\ }\href {\doibase 10.1103/PhysRevLett.110.168703}
  {\bibfield  {journal} {\bibinfo  {journal} {Phys. Rev. Lett.}\ }\textbf
  {\bibinfo {volume} {110}},\ \bibinfo {pages} {168703} (\bibinfo {year}
  {2013})}\BibitemShut {NoStop}%
\bibitem [{\citenamefont {Villegas}\ \emph {et~al.}(2023)\citenamefont
  {Villegas}, \citenamefont {Gili}, \citenamefont {Caldarelli},\ and\
  \citenamefont {Gabrielli}}]{LRG}%
  \BibitemOpen
  \bibfield  {author} {\bibinfo {author} {\bibfnamefont {P.}~\bibnamefont
  {Villegas}}, \bibinfo {author} {\bibfnamefont {T.}~\bibnamefont {Gili}},
  \bibinfo {author} {\bibfnamefont {G.}~\bibnamefont {Caldarelli}}, \ and\
  \bibinfo {author} {\bibfnamefont {A.}~\bibnamefont {Gabrielli}},\ }\href
  {\doibase 10.1038/s41567-022-01866-8} {\bibfield  {journal} {\bibinfo
  {journal} {Nat. Phys.}\ }\textbf {\bibinfo {volume} {19}},\ \bibinfo {pages}
  {445} (\bibinfo {year} {2023})}\BibitemShut {NoStop}%
\bibitem [{\citenamefont {Burioni}\ and\ \citenamefont
  {Cassi}(1996)}]{Cassi1996}%
  \BibitemOpen
  \bibfield  {author} {\bibinfo {author} {\bibfnamefont {R.}~\bibnamefont
  {Burioni}}\ and\ \bibinfo {author} {\bibfnamefont {D.}~\bibnamefont
  {Cassi}},\ }\href {\doibase 10.1103/PhysRevLett.76.1091} {\bibfield
  {journal} {\bibinfo  {journal} {Phys. Rev. Lett.}\ }\textbf {\bibinfo
  {volume} {76}},\ \bibinfo {pages} {1091} (\bibinfo {year}
  {1996})}\BibitemShut {NoStop}%
\bibitem [{\citenamefont {Cassi}(1992)}]{Cassi1992}%
  \BibitemOpen
  \bibfield  {author} {\bibinfo {author} {\bibfnamefont {D.}~\bibnamefont
  {Cassi}},\ }\href {\doibase 10.1103/PhysRevLett.68.3631} {\bibfield
  {journal} {\bibinfo  {journal} {Phys. Rev. Lett.}\ }\textbf {\bibinfo
  {volume} {68}},\ \bibinfo {pages} {3631} (\bibinfo {year}
  {1992})}\BibitemShut {NoStop}%
\bibitem [{\citenamefont {Villegas}\ \emph {et~al.}(2025)\citenamefont
  {Villegas}, \citenamefont {Gabrielli}, \citenamefont {Poggialini},\ and\
  \citenamefont {Gili}}]{Modularity}%
  \BibitemOpen
  \bibfield  {author} {\bibinfo {author} {\bibfnamefont {P.}~\bibnamefont
  {Villegas}}, \bibinfo {author} {\bibfnamefont {A.}~\bibnamefont {Gabrielli}},
  \bibinfo {author} {\bibfnamefont {A.}~\bibnamefont {Poggialini}}, \ and\
  \bibinfo {author} {\bibfnamefont {T.}~\bibnamefont {Gili}},\ }\href {\doibase
  10.1103/PhysRevResearch.7.013065} {\bibfield  {journal} {\bibinfo  {journal}
  {Phys. Rev. Res.}\ }\textbf {\bibinfo {volume} {7}},\ \bibinfo {pages}
  {013065} (\bibinfo {year} {2025})}\BibitemShut {NoStop}%
\bibitem [{\citenamefont {Gu}\ \emph {et~al.}(2021)\citenamefont {Gu},
  \citenamefont {Tandon}, \citenamefont {Ahn} \emph {et~al.}}]{Gu2021}%
  \BibitemOpen
  \bibfield  {author} {\bibinfo {author} {\bibfnamefont {W.}~\bibnamefont
  {Gu}}, \bibinfo {author} {\bibfnamefont {A.}~\bibnamefont {Tandon}}, \bibinfo
  {author} {\bibfnamefont {Y.-Y.}\ \bibnamefont {Ahn}},  \emph {et~al.},\
  }\href {\doibase 10.1038/s41467-021-23795-5} {\bibfield  {journal} {\bibinfo
  {journal} {Nat. Comm.}\ }\textbf {\bibinfo {volume} {12}},\ \bibinfo {pages}
  {3772} (\bibinfo {year} {2021})}\BibitemShut {NoStop}%
\bibitem [{\citenamefont {Bogu{\~n}{\'a}}\ \emph {et~al.}(2021)\citenamefont
  {Bogu{\~n}{\'a}}, \citenamefont {Bonamassa}, \citenamefont {Domenico} \emph
  {et~al.}}]{Boguna2021}%
  \BibitemOpen
  \bibfield  {author} {\bibinfo {author} {\bibfnamefont {M.}~\bibnamefont
  {Bogu{\~n}{\'a}}}, \bibinfo {author} {\bibfnamefont {I.}~\bibnamefont
  {Bonamassa}}, \bibinfo {author} {\bibfnamefont {M.~D.}\ \bibnamefont
  {Domenico}},  \emph {et~al.},\ }\href {\doibase 10.1038/s42254-020-00264-4}
  {\bibfield  {journal} {\bibinfo  {journal} {Nat. Rev. Phys.}\ }\textbf
  {\bibinfo {volume} {3}},\ \bibinfo {pages} {114} (\bibinfo {year}
  {2021})}\BibitemShut {NoStop}%
\bibitem [{\citenamefont {Serrano}\ \emph {et~al.}(2008)\citenamefont
  {Serrano}, \citenamefont {Krioukov},\ and\ \citenamefont
  {Bogu\~n\'a}}]{Hyper1}%
  \BibitemOpen
  \bibfield  {author} {\bibinfo {author} {\bibfnamefont {M.~A.}\ \bibnamefont
  {Serrano}}, \bibinfo {author} {\bibfnamefont {D.}~\bibnamefont {Krioukov}}, \
  and\ \bibinfo {author} {\bibfnamefont {M.}~\bibnamefont {Bogu\~n\'a}},\
  }\href {\doibase 10.1103/PhysRevLett.100.078701} {\bibfield  {journal}
  {\bibinfo  {journal} {Phys. Rev. Lett.}\ }\textbf {\bibinfo {volume} {100}},\
  \bibinfo {pages} {078701} (\bibinfo {year} {2008})}\BibitemShut {NoStop}%
\bibitem [{\citenamefont {Krioukov}\ \emph {et~al.}(2010)\citenamefont
  {Krioukov}, \citenamefont {Papadopoulos}, \citenamefont {Kitsak},
  \citenamefont {Vahdat},\ and\ \citenamefont {Bogu\~n\'a}}]{Hyper2}%
  \BibitemOpen
  \bibfield  {author} {\bibinfo {author} {\bibfnamefont {D.}~\bibnamefont
  {Krioukov}}, \bibinfo {author} {\bibfnamefont {F.}~\bibnamefont
  {Papadopoulos}}, \bibinfo {author} {\bibfnamefont {M.}~\bibnamefont
  {Kitsak}}, \bibinfo {author} {\bibfnamefont {A.}~\bibnamefont {Vahdat}}, \
  and\ \bibinfo {author} {\bibfnamefont {M.}~\bibnamefont {Bogu\~n\'a}},\
  }\href {\doibase 10.1103/PhysRevE.82.036106} {\bibfield  {journal} {\bibinfo
  {journal} {Phys. Rev. E}\ }\textbf {\bibinfo {volume} {82}},\ \bibinfo
  {pages} {036106} (\bibinfo {year} {2010})}\BibitemShut {NoStop}%
\bibitem [{\citenamefont {Bogu{\~n}{\'a}}\ \emph {et~al.}(2009)\citenamefont
  {Bogu{\~n}{\'a}}, \citenamefont {Krioukov},\ and\ \citenamefont
  {Claffy}}]{Boguna2009}%
  \BibitemOpen
  \bibfield  {author} {\bibinfo {author} {\bibfnamefont {M.}~\bibnamefont
  {Bogu{\~n}{\'a}}}, \bibinfo {author} {\bibfnamefont {D.}~\bibnamefont
  {Krioukov}}, \ and\ \bibinfo {author} {\bibfnamefont {K.~C.}\ \bibnamefont
  {Claffy}},\ }\href {\doibase 10.1038/nphys1150} {\bibfield  {journal}
  {\bibinfo  {journal} {Nat. Phys.}\ }\textbf {\bibinfo {volume} {5}},\
  \bibinfo {pages} {74} (\bibinfo {year} {2009})}\BibitemShut {NoStop}%
\bibitem [{\citenamefont {Guly{\'a}s}\ \emph {et~al.}(2015)\citenamefont
  {Guly{\'a}s}, \citenamefont {B{\'i}r{\'o}}, \citenamefont {K{\H{o}}r{\"o}si},
  \citenamefont {R{\'e}tv{\'a}ri},\ and\ \citenamefont
  {Krioukov}}]{Gulyas2015}%
  \BibitemOpen
  \bibfield  {author} {\bibinfo {author} {\bibfnamefont {A.}~\bibnamefont
  {Guly{\'a}s}}, \bibinfo {author} {\bibfnamefont {J.~J.}\ \bibnamefont
  {B{\'i}r{\'o}}}, \bibinfo {author} {\bibfnamefont {A.}~\bibnamefont
  {K{\H{o}}r{\"o}si}}, \bibinfo {author} {\bibfnamefont {G.}~\bibnamefont
  {R{\'e}tv{\'a}ri}}, \ and\ \bibinfo {author} {\bibfnamefont {D.}~\bibnamefont
  {Krioukov}},\ }\href {\doibase 10.1038/ncomms8651} {\bibfield  {journal}
  {\bibinfo  {journal} {Nat. Comm.}\ }\textbf {\bibinfo {volume} {6}},\
  \bibinfo {pages} {7651} (\bibinfo {year} {2015})}\BibitemShut {NoStop}%
\bibitem [{\citenamefont {Almagro}\ \emph {et~al.}(2022)\citenamefont
  {Almagro}, \citenamefont {Bogu\~n\'a},\ and\ \citenamefont
  {Serrano}}]{Almagro2022}%
  \BibitemOpen
  \bibfield  {author} {\bibinfo {author} {\bibfnamefont {P.}~\bibnamefont
  {Almagro}}, \bibinfo {author} {\bibfnamefont {M.}~\bibnamefont {Bogu\~n\'a}},
  \ and\ \bibinfo {author} {\bibfnamefont {M.~A.}\ \bibnamefont {Serrano}},\
  }\href {\doibase 10.1038/s41467-022-33685-z} {\bibfield  {journal} {\bibinfo
  {journal} {Nat. Comm.}\ }\textbf {\bibinfo {volume} {13}},\ \bibinfo {pages}
  {6096} (\bibinfo {year} {2022})}\BibitemShut {NoStop}%
\bibitem [{\citenamefont {Kardar}(2007)}]{KardarBook}%
  \BibitemOpen
  \bibfield  {author} {\bibinfo {author} {\bibfnamefont {M.}~\bibnamefont
  {Kardar}},\ }\href {\doibase 10.1017/CBO9780511815881} {\emph {\bibinfo
  {title} {Statistical physics of fields}}}\ (\bibinfo  {publisher} {Cambridge
  University Press},\ \bibinfo {address} {Cambridge},\ \bibinfo {year}
  {2007})\BibitemShut {NoStop}%
\bibitem [{\citenamefont {Belkin}\ and\ \citenamefont {Niyogi}(2001)}]{Belkin}%
  \BibitemOpen
  \bibfield  {author} {\bibinfo {author} {\bibfnamefont {M.}~\bibnamefont
  {Belkin}}\ and\ \bibinfo {author} {\bibfnamefont {P.}~\bibnamefont
  {Niyogi}},\ }in\ \href@noop {} {\emph {\bibinfo {booktitle} {Proceedings of
  the 14th International Conference on Neural Information Processing Systems:
  Natural and Synthetic (NIPS'01)}}}\ (\bibinfo  {publisher} {MIT Press},\
  \bibinfo {address} {Cambridge, MA, USA},\ \bibinfo {year} {2001})\ pp.\
  \bibinfo {pages} {585--591}\BibitemShut {NoStop}%
\bibitem [{SM()}]{SM}%
  \BibitemOpen
  \href@noop {} {}\bibinfo {note} {See Supplemental Material at [] for further
  technical details, extended results, and additional simulations supporting
  the findings in the main text.}\BibitemShut {Stop}%
\bibitem [{\citenamefont {Holme}\ and\ \citenamefont {Kim}(2002)}]{KimHolme}%
  \BibitemOpen
  \bibfield  {author} {\bibinfo {author} {\bibfnamefont {P.}~\bibnamefont
  {Holme}}\ and\ \bibinfo {author} {\bibfnamefont {B.~J.}\ \bibnamefont
  {Kim}},\ }\href {\doibase 10.1103/PhysRevE.65.026107} {\bibfield  {journal}
  {\bibinfo  {journal} {Phys. Rev. E}\ }\textbf {\bibinfo {volume} {65}},\
  \bibinfo {pages} {026107} (\bibinfo {year} {2002})}\BibitemShut {NoStop}%
\bibitem [{\citenamefont {Moretti}\ and\ \citenamefont
  {Mu{\~n}oz}(2013)}]{moretti2013griffiths}%
  \BibitemOpen
  \bibfield  {author} {\bibinfo {author} {\bibfnamefont {P.}~\bibnamefont
  {Moretti}}\ and\ \bibinfo {author} {\bibfnamefont {M.~A.}\ \bibnamefont
  {Mu{\~n}oz}},\ }\href {\doibase 10.1038/ncomms3521} {\bibfield  {journal}
  {\bibinfo  {journal} {Nat. Comm.}\ }\textbf {\bibinfo {volume} {4}},\
  \bibinfo {pages} {2521} (\bibinfo {year} {2013})}\BibitemShut {NoStop}%
\bibitem [{\citenamefont {Villegas}\ \emph {et~al.}(2014)\citenamefont
  {Villegas}, \citenamefont {Moretti},\ and\ \citenamefont
  {Mu{\~n}oz}}]{SciRep}%
  \BibitemOpen
  \bibfield  {author} {\bibinfo {author} {\bibfnamefont {P.}~\bibnamefont
  {Villegas}}, \bibinfo {author} {\bibfnamefont {P.}~\bibnamefont {Moretti}}, \
  and\ \bibinfo {author} {\bibfnamefont {M.~A.}\ \bibnamefont {Mu{\~n}oz}},\
  }\href {\doibase 10.1038/srep05990} {\bibfield  {journal} {\bibinfo
  {journal} {Sci. Rep.}\ }\textbf {\bibinfo {volume} {4}},\ \bibinfo {pages}
  {5990} (\bibinfo {year} {2014})}\BibitemShut {NoStop}%
\bibitem [{\citenamefont {Dorogovtsev}\ \emph {et~al.}(2002)\citenamefont
  {Dorogovtsev}, \citenamefont {Goltsev},\ and\ \citenamefont {Mendes}}]{DGM}%
  \BibitemOpen
  \bibfield  {author} {\bibinfo {author} {\bibfnamefont {S.~N.}\ \bibnamefont
  {Dorogovtsev}}, \bibinfo {author} {\bibfnamefont {A.~V.}\ \bibnamefont
  {Goltsev}}, \ and\ \bibinfo {author} {\bibfnamefont {J.~F.~F.}\ \bibnamefont
  {Mendes}},\ }\href {\doibase 10.1103/PhysRevE.65.066122} {\bibfield
  {journal} {\bibinfo  {journal} {Phys. Rev. E}\ }\textbf {\bibinfo {volume}
  {65}},\ \bibinfo {pages} {066122} (\bibinfo {year} {2002})}\BibitemShut
  {NoStop}%
\bibitem [{\citenamefont {Kittel}\ and\ \citenamefont
  {McEuen}(2005)}]{Kittel2005}%
  \BibitemOpen
  \bibfield  {author} {\bibinfo {author} {\bibfnamefont {C.}~\bibnamefont
  {Kittel}}\ and\ \bibinfo {author} {\bibfnamefont {P.}~\bibnamefont
  {McEuen}},\ }\href@noop {} {\emph {\bibinfo {title} {Introduction to Solid
  State Physics}}},\ \bibinfo {edition} {8th}\ ed.\ (\bibinfo  {publisher}
  {Wiley},\ \bibinfo {address} {New York},\ \bibinfo {year} {2005})\BibitemShut
  {NoStop}%
\bibitem [{\citenamefont {Xin}\ \emph {et~al.}(2024)\citenamefont {Xin},
  \citenamefont {Falsi}, \citenamefont {Gelkop}, \citenamefont {Pierangeli},
  \citenamefont {Zhang}, \citenamefont {Bo}, \citenamefont {Fusella},
  \citenamefont {Agranat},\ and\ \citenamefont {DelRe}}]{Xin2024}%
  \BibitemOpen
  \bibfield  {author} {\bibinfo {author} {\bibfnamefont {F.}~\bibnamefont
  {Xin}}, \bibinfo {author} {\bibfnamefont {L.}~\bibnamefont {Falsi}}, \bibinfo
  {author} {\bibfnamefont {Y.}~\bibnamefont {Gelkop}}, \bibinfo {author}
  {\bibfnamefont {D.}~\bibnamefont {Pierangeli}}, \bibinfo {author}
  {\bibfnamefont {G.}~\bibnamefont {Zhang}}, \bibinfo {author} {\bibfnamefont
  {F.}~\bibnamefont {Bo}}, \bibinfo {author} {\bibfnamefont {F.}~\bibnamefont
  {Fusella}}, \bibinfo {author} {\bibfnamefont {A.~J.}\ \bibnamefont
  {Agranat}}, \ and\ \bibinfo {author} {\bibfnamefont {E.}~\bibnamefont
  {DelRe}},\ }\href {\doibase 10.1103/PhysRevLett.132.066603} {\bibfield
  {journal} {\bibinfo  {journal} {Phys. Rev. Lett.}\ }\textbf {\bibinfo
  {volume} {132}},\ \bibinfo {pages} {066603} (\bibinfo {year}
  {2024})}\BibitemShut {NoStop}%
\bibitem [{\citenamefont {Grassberger}\ and\ \citenamefont
  {Procaccia}(1983)}]{CorrDim}%
  \BibitemOpen
  \bibfield  {author} {\bibinfo {author} {\bibfnamefont {P.}~\bibnamefont
  {Grassberger}}\ and\ \bibinfo {author} {\bibfnamefont {I.}~\bibnamefont
  {Procaccia}},\ }\href {\doibase 10.1016/0167-2789(83)90298-1} {\bibfield
  {journal} {\bibinfo  {journal} {Phys. D: Nonlinear Phenom.}\ }\textbf
  {\bibinfo {volume} {9}},\ \bibinfo {pages} {189–208} (\bibinfo {year}
  {1983})}\BibitemShut {NoStop}%
\bibitem [{\citenamefont {Villegas}\ \emph {et~al.}(2024)\citenamefont
  {Villegas}, \citenamefont {Gili}, \citenamefont {Caldarelli},\ and\
  \citenamefont {Gabrielli}}]{ScaleFree}%
  \BibitemOpen
  \bibfield  {author} {\bibinfo {author} {\bibfnamefont {P.}~\bibnamefont
  {Villegas}}, \bibinfo {author} {\bibfnamefont {T.}~\bibnamefont {Gili}},
  \bibinfo {author} {\bibfnamefont {G.}~\bibnamefont {Caldarelli}}, \ and\
  \bibinfo {author} {\bibfnamefont {A.}~\bibnamefont {Gabrielli}},\ }\href
  {\doibase 10.1103/PhysRevE.109.L042402} {\bibfield  {journal} {\bibinfo
  {journal} {Phys. Rev. E}\ }\textbf {\bibinfo {volume} {109}},\ \bibinfo
  {pages} {L042402} (\bibinfo {year} {2024})}\BibitemShut {NoStop}%
\bibitem [{\citenamefont {Falconer}(2014)}]{Falconer2014}%
  \BibitemOpen
  \bibfield  {author} {\bibinfo {author} {\bibfnamefont {K.}~\bibnamefont
  {Falconer}},\ }\href@noop {} {\emph {\bibinfo {title} {Fractal Geometry:
  Mathematical Foundations and Applications}}},\ \bibinfo {edition} {3rd}\ ed.\
  (\bibinfo  {publisher} {John Wiley \& Sons},\ \bibinfo {address} {New York},\
  \bibinfo {year} {2014})\BibitemShut {NoStop}%
\bibitem [{\citenamefont {Rams}\ and\ \citenamefont {Simon}(2014)}]{Rams2014}%
  \BibitemOpen
  \bibfield  {author} {\bibinfo {author} {\bibfnamefont {M.}~\bibnamefont
  {Rams}}\ and\ \bibinfo {author} {\bibfnamefont {K.}~\bibnamefont {Simon}},\
  }in\ \href@noop {} {\emph {\bibinfo {booktitle} {Geometry and Analysis of
  Fractals}}},\ \bibinfo {series} {Springer Proceedings in Mathematics \&
  Statistics}, Vol.~\bibinfo {volume} {88},\ \bibinfo {editor} {edited by\
  \bibinfo {editor} {\bibfnamefont {D.~J.}\ \bibnamefont {Feng}}\ and\ \bibinfo
  {editor} {\bibfnamefont {K.~S.}\ \bibnamefont {Lau}}}\ (\bibinfo  {publisher}
  {Springer},\ \bibinfo {address} {New York},\ \bibinfo {year}
  {2014})\BibitemShut {NoStop}%
\bibitem [{\citenamefont {Falsi}\ \emph {et~al.}(2021)\citenamefont {Falsi},
  \citenamefont {Aversa}, \citenamefont {Di~Mei}, \citenamefont {Pierangeli},
  \citenamefont {Xin}, \citenamefont {Agranat},\ and\ \citenamefont
  {DelRe}}]{Falsi2021}%
  \BibitemOpen
  \bibfield  {author} {\bibinfo {author} {\bibfnamefont {L.}~\bibnamefont
  {Falsi}}, \bibinfo {author} {\bibfnamefont {M.}~\bibnamefont {Aversa}},
  \bibinfo {author} {\bibfnamefont {F.}~\bibnamefont {Di~Mei}}, \bibinfo
  {author} {\bibfnamefont {D.}~\bibnamefont {Pierangeli}}, \bibinfo {author}
  {\bibfnamefont {F.}~\bibnamefont {Xin}}, \bibinfo {author} {\bibfnamefont
  {A.~J.}\ \bibnamefont {Agranat}}, \ and\ \bibinfo {author} {\bibfnamefont
  {E.}~\bibnamefont {DelRe}},\ }\href {\doibase 10.1103/PhysRevLett.126.037601}
  {\bibfield  {journal} {\bibinfo  {journal} {Phys. Rev. Lett.}\ }\textbf
  {\bibinfo {volume} {126}},\ \bibinfo {pages} {037601} (\bibinfo {year}
  {2021})}\BibitemShut {NoStop}%
\bibitem [{\citenamefont {Eckmann}\ and\ \citenamefont
  {Ruelle}(1985)}]{Eckmann1985}%
  \BibitemOpen
  \bibfield  {author} {\bibinfo {author} {\bibfnamefont {J.~P.}\ \bibnamefont
  {Eckmann}}\ and\ \bibinfo {author} {\bibfnamefont {D.}~\bibnamefont
  {Ruelle}},\ }\href {\doibase 10.1103/RevModPhys.57.617} {\bibfield  {journal}
  {\bibinfo  {journal} {Rev. Mod. Phys.}\ }\textbf {\bibinfo {volume} {57}},\
  \bibinfo {pages} {617} (\bibinfo {year} {1985})}\BibitemShut {NoStop}%
\bibitem [{\citenamefont {Argyris}\ \emph {et~al.}(1998)\citenamefont
  {Argyris}, \citenamefont {Andreadis}, \citenamefont {Pavlos},\ and\
  \citenamefont {Athanasiou}}]{Argyrys1998}%
  \BibitemOpen
  \bibfield  {author} {\bibinfo {author} {\bibfnamefont {J.}~\bibnamefont
  {Argyris}}, \bibinfo {author} {\bibfnamefont {I.}~\bibnamefont {Andreadis}},
  \bibinfo {author} {\bibfnamefont {G.}~\bibnamefont {Pavlos}}, \ and\ \bibinfo
  {author} {\bibfnamefont {M.}~\bibnamefont {Athanasiou}},\ }\href {\doibase
  10.1016/S0960-0779(97)00120-3} {\bibfield  {journal} {\bibinfo  {journal}
  {Chaos Solit. Fractals}\ }\textbf {\bibinfo {volume} {9}},\ \bibinfo {pages}
  {343–361} (\bibinfo {year} {1998})}\BibitemShut {NoStop}%
\bibitem [{\citenamefont {Aref}\ and\ \citenamefont {Wilson}(2019)}]{Aref2019}%
  \BibitemOpen
  \bibfield  {author} {\bibinfo {author} {\bibfnamefont {S.}~\bibnamefont
  {Aref}}\ and\ \bibinfo {author} {\bibfnamefont {M.~C.}\ \bibnamefont
  {Wilson}},\ }\href {\doibase 10.1093/comnet/cny015} {\bibfield  {journal}
  {\bibinfo  {journal} {J. Complex Netw.}\ }\textbf {\bibinfo {volume} {7}},\
  \bibinfo {pages} {163} (\bibinfo {year} {2019})}\BibitemShut {NoStop}%
\bibitem [{\citenamefont {Lu}\ and\ \citenamefont {Tang}(2004)}]{Lu2004}%
  \BibitemOpen
  \bibfield  {author} {\bibinfo {author} {\bibfnamefont {Y.}~\bibnamefont
  {Lu}}\ and\ \bibinfo {author} {\bibfnamefont {J.}~\bibnamefont {Tang}},\
  }\href {\doibase 10.1068/b3163} {\bibfield  {journal} {\bibinfo  {journal}
  {Environ. Plan. B: Plan. Des.}\ }\textbf {\bibinfo {volume} {31}},\ \bibinfo
  {pages} {895} (\bibinfo {year} {2004})}\BibitemShut {NoStop}%
\bibitem [{\citenamefont {Dallas}(2016)}]{Dallas2016}%
  \BibitemOpen
  \bibfield  {author} {\bibinfo {author} {\bibfnamefont {T.}~\bibnamefont
  {Dallas}},\ }\href {\doibase 10.1111/ecog.02131} {\bibfield  {journal}
  {\bibinfo  {journal} {Ecography}\ }\textbf {\bibinfo {volume} {39}},\
  \bibinfo {pages} {391–393} (\bibinfo {year} {2016})}\BibitemShut {NoStop}%
\bibitem [{\citenamefont {Dallas}\ \emph {et~al.}(2018)\citenamefont {Dallas},
  \citenamefont {Aguirre}, \citenamefont {Budischak}, \citenamefont {Carlson},
  \citenamefont {Ezenwa}, \citenamefont {Han}, \citenamefont {Huang},\ and\
  \citenamefont {Stephens}}]{Dallas2018}%
  \BibitemOpen
  \bibfield  {author} {\bibinfo {author} {\bibfnamefont {T.~A.}\ \bibnamefont
  {Dallas}}, \bibinfo {author} {\bibfnamefont {A.~A.}\ \bibnamefont {Aguirre}},
  \bibinfo {author} {\bibfnamefont {S.}~\bibnamefont {Budischak}}, \bibinfo
  {author} {\bibfnamefont {C.}~\bibnamefont {Carlson}}, \bibinfo {author}
  {\bibfnamefont {V.}~\bibnamefont {Ezenwa}}, \bibinfo {author} {\bibfnamefont
  {B.}~\bibnamefont {Han}}, \bibinfo {author} {\bibfnamefont {S.}~\bibnamefont
  {Huang}}, \ and\ \bibinfo {author} {\bibfnamefont {P.~R.}\ \bibnamefont
  {Stephens}},\ }\href {\doibase 10.1111/geb.12829} {\bibfield  {journal}
  {\bibinfo  {journal} {Glob. Ecol. Biogeogr.}\ }\textbf {\bibinfo {volume}
  {27}},\ \bibinfo {pages} {1437} (\bibinfo {year} {2018})}\BibitemShut
  {NoStop}%
\bibitem [{\citenamefont {Simonis}\ \emph {et~al.}(2009)\citenamefont {Simonis}
  \emph {et~al.}}]{Simonis2009}%
  \BibitemOpen
  \bibfield  {author} {\bibinfo {author} {\bibfnamefont {N.}~\bibnamefont
  {Simonis}} \emph {et~al.},\ }\href {\doibase 10.1038/nmeth.1280} {\bibfield
  {journal} {\bibinfo  {journal} {Nat. Methods}\ }\textbf {\bibinfo {volume}
  {6}},\ \bibinfo {pages} {47} (\bibinfo {year} {2009})}\BibitemShut {NoStop}%
\bibitem [{\citenamefont {Gunsalus}\ \emph {et~al.}(2005)\citenamefont
  {Gunsalus} \emph {et~al.}}]{Gunsalus2005}%
  \BibitemOpen
  \bibfield  {author} {\bibinfo {author} {\bibfnamefont {K.~C.}\ \bibnamefont
  {Gunsalus}} \emph {et~al.},\ }\href {\doibase 10.1038/nature03885} {\bibfield
   {journal} {\bibinfo  {journal} {Nature}\ }\textbf {\bibinfo {volume}
  {436}},\ \bibinfo {pages} {861} (\bibinfo {year} {2005})}\BibitemShut
  {NoStop}%
\bibitem [{\citenamefont {Donetti}\ and\ \citenamefont
  {Mu{\~n}oz}(2004)}]{Donetti}%
  \BibitemOpen
  \bibfield  {author} {\bibinfo {author} {\bibfnamefont {L.}~\bibnamefont
  {Donetti}}\ and\ \bibinfo {author} {\bibfnamefont {M.~A.}\ \bibnamefont
  {Mu{\~n}oz}},\ }\href {\doibase 10.1088/1742-5468/2004/10/P10012} {\bibfield
  {journal} {\bibinfo  {journal} {J. Stat. Mech.: Theory Exp.}\ }\textbf
  {\bibinfo {volume} {2004}},\ \bibinfo {pages} {P10012} (\bibinfo {year}
  {2004})}\BibitemShut {NoStop}%
\bibitem [{\citenamefont {Crutchfield}(2012)}]{Crut2012}%
  \BibitemOpen
  \bibfield  {author} {\bibinfo {author} {\bibfnamefont {J.~P.}\ \bibnamefont
  {Crutchfield}},\ }\href {\doibase 10.1038/nphys2190} {\bibfield  {journal}
  {\bibinfo  {journal} {Nat. Phys.}\ }\textbf {\bibinfo {volume} {8}},\
  \bibinfo {pages} {17} (\bibinfo {year} {2012})}\BibitemShut {NoStop}%
\bibitem [{\citenamefont {Strogatz}(2001)}]{Strogatz2001}%
  \BibitemOpen
  \bibfield  {author} {\bibinfo {author} {\bibfnamefont {S.~H.}\ \bibnamefont
  {Strogatz}},\ }\href {\doibase 10.1038/35065725} {\bibfield  {journal}
  {\bibinfo  {journal} {Nature}\ }\textbf {\bibinfo {volume} {410}},\ \bibinfo
  {pages} {268} (\bibinfo {year} {2001})}\BibitemShut {NoStop}%
\bibitem [{\citenamefont {Dorogovtsev}\ \emph {et~al.}(2008)\citenamefont
  {Dorogovtsev}, \citenamefont {Goltsev},\ and\ \citenamefont
  {Mendes}}]{Doro2008}%
  \BibitemOpen
  \bibfield  {author} {\bibinfo {author} {\bibfnamefont {S.~N.}\ \bibnamefont
  {Dorogovtsev}}, \bibinfo {author} {\bibfnamefont {A.~V.}\ \bibnamefont
  {Goltsev}}, \ and\ \bibinfo {author} {\bibfnamefont {J.~F.~F.}\ \bibnamefont
  {Mendes}},\ }\href {\doibase 10.1103/RevModPhys.80.1275} {\bibfield
  {journal} {\bibinfo  {journal} {Rev. Mod. Phys.}\ }\textbf {\bibinfo {volume}
  {80}},\ \bibinfo {pages} {1275} (\bibinfo {year} {2008})}\BibitemShut
  {NoStop}%
\bibitem [{\citenamefont {Barth{\'e}lemy}(2011)}]{Barth2011}%
  \BibitemOpen
  \bibfield  {author} {\bibinfo {author} {\bibfnamefont {M.}~\bibnamefont
  {Barth{\'e}lemy}},\ }\href {\doibase 10.1016/j.physrep.2010.11.002}
  {\bibfield  {journal} {\bibinfo  {journal} {Phys. Rep.}\ }\textbf {\bibinfo
  {volume} {499}},\ \bibinfo {pages} {1} (\bibinfo {year} {2011})}\BibitemShut
  {NoStop}%
\bibitem [{\citenamefont {Peixoto}(2024)}]{Peixoto2024}%
  \BibitemOpen
  \bibfield  {author} {\bibinfo {author} {\bibfnamefont {T.~P.}\ \bibnamefont
  {Peixoto}},\ }\href {\doibase 10.5281/zenodo.7839981} {\enquote {\bibinfo
  {title} {The netzschleuder network catalogue and repository}}\ } (\bibinfo
  {year} {2024})\BibitemShut {NoStop}%
\bibitem [{GH()}]{GH}%
  \BibitemOpen
  \href {https://github.com/pvgongora/NetAttractors/} {}\bibinfo {note}
  {\href{https://github.com/pvgongora/NetAttractors/}{https://github.com/pvgongora/NetAttractors/}}\BibitemShut
  {NoStop}%
\end{thebibliography}

\clearpage


\section*{Supplementary material}

# Supplemental Material: Strange Attractors in Complex Networks

Pablo Villegas[1, 2, *]

[1] *'Enrico Fermi' Research Center (CREF), Via Panisperna 89A, 00184 - Rome, Italy*
[2] *Instituto Carlos I de Física Teórica y Computacional, Univ. de Granada, E-18071, Granada, Spain.*

## CONTENTS

---

* pablo.villegas@cref.it

## 1. SCALE-INVARIANT NETWORKS.

**Kim and Holme networks** The procedure proposed by Kim and Holme (KH) works by adding a new step to the preferential attachment rule [1]. In particular, the procedure is as follows:

(i) A vertex with m edges is added at each time step. The growth time $t$ corresponds to the number of time steps.

(ii) Preferential attachment rule (PA): Each edge of $v$ is then linked to an already existing vertex with probability proportional to its degree, namely, $P_i = \frac{\kappa_i}{\sum_j \kappa_j}$.

(iii) Triad formation (TF): After each PA step in which a new vertex v is added and some edge $(u, v)$ is added, a triangle is closed with probability $p$ by choosing a neighbor of $u$, $u2$ and adding the edge $(v, u2)$.

**Hierarchical Modular Networks.** A HMN consists of N nodes and L links, organized into hierarchical levels for easy analysis and built using a bottom-to-top approach. First, local fully connected modules are built, grouping them recursively by establishing new inter-modular links in a deterministic manner with a level-dependent number of connections (HMN). Hence, the growing algorithm works as follows, as detailed in [2]:

(i) At each hierarchical level $l = 1, 2, ..., s$, different pairs of blocks are selected, each with size $2^{i-1}M_0$. All possible undirected $4^{i-1}M_0^2$ connections between the two blocks are evaluated and established to avoid repetitions.

(ii) The number of connections between blocks at each level is set a priori at a constant value $\alpha$.

(iii) The method is stochastic in assigning connections, although the number of them (as well as the degree of the network) is fixed deterministically, being,

$$\langle \kappa \rangle = M_0 - 1 + \frac{2\alpha}{M_0}(1 - 2^{-s}).$$

**DGM networks.** These are deterministic and recursive networks [3]. The growth begins with a single edge connecting two vertices at $t = -1$. At each subsequent time step, a new vertex is added to every edge of the graph, where the new vertex is connected to both of the edge's endpoints. At $t = 0$, this results in a triangle of edges linking three vertices. By $t = 1$, the graph consists of 6 vertices and nine edges, and the structure continues to grow. The total number of vertices at time $t$ is given by $N_t = \frac{3(3^t+1)}{2}$, and the total number of edges is $L_t = 3^{t+1}$, leading to an average degree of $\langle \kappa \rangle = \frac{2L_t}{N_t} = \frac{4}{1+3^{-t}}$.

## 2. 2D PROJECTIONS

Figure 1 reports the 2D projections using the two first eigenvalues, $\lambda_1$ and $\lambda_2$, for the different networks considered in the first figure of the main text. Note that these planar projections are the analog of far-field images obtained for disordered materials [4, 5].

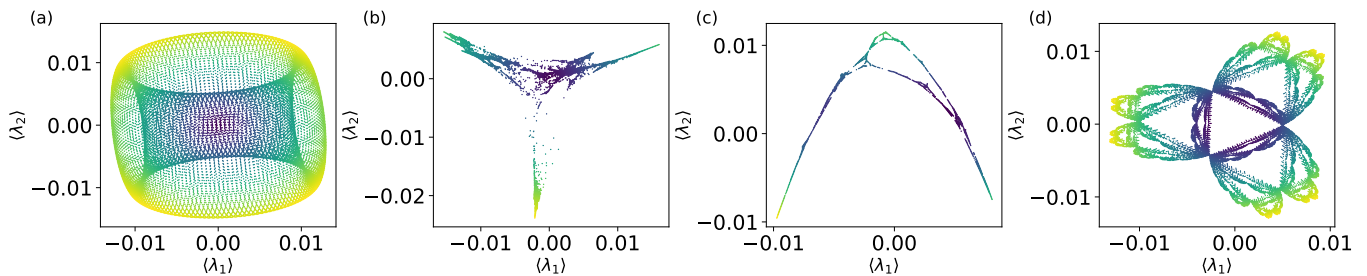

FIG. 1. **2D embeddings.** Low-dimensional representation using the two smallest low-frequency normalized network modes, $|\hat{\lambda}_1\rangle$, and $|\hat{\lambda}_2\rangle$ as the coordinate axes for: **(a)** A 2D lattice of size $L = 128$. **(b)** A KH network of size $N = 2^{14}$ nodes with $m = 3$ and $p = 1$. **(c)** A HMN network with $s = 13, \alpha = 3,$, and $M_0 = 3$ consisting on $N = 24576$ nodes. **(d)** A DGM network of size $N = 29526$.

## 3. CORRELATION INTEGRAL ANALYSIS

### Embedding dimension analysis

The correlation integral analysis as a function of the system size has been reported in the main text. Here, the correlation integral for the different synthetic networks analyzed in the main text and large system sizes is shown in Figure 2. Note that all networks present a well-defined scaling region for the condition $D > d$.

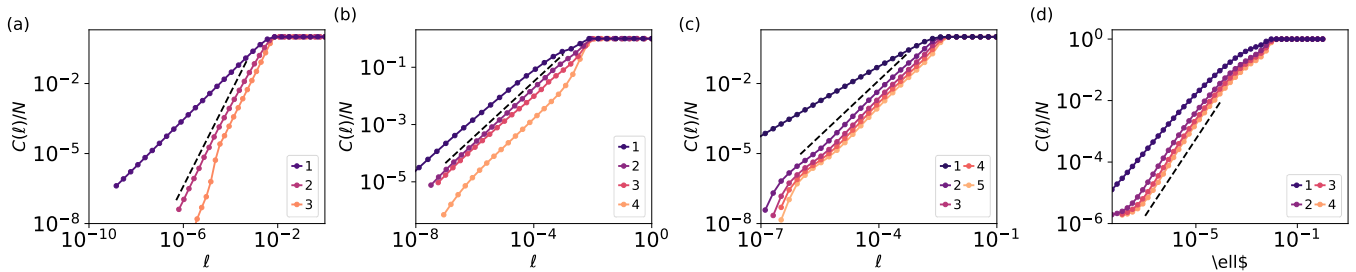

FIG. 2. **d-dimensional embeddings.** Correlation integral, $C(\ell)$ versus distance, $\ell$, for different values of $d$ (see legends). Black dashed lines represent the scaling as predicted by half of the network spectral dimension, $d_s/2$, or the Euclidean dimension $d_E$ for the case of the 2D lattice. **(a)** 2D lattice of size $N = 2^{18}$. **(b)** HMN networks with $\alpha = 3$, and $M_0 = 3$, of size $N = M_0 2^s$ with $s = 16$. **(c)** DGM networks of size $N = {}^{3+3^s}/_2$ with $s = 13$. **(b)** KH networks with $m = 2$ and $p = 1$ of size $N = 2^s$ and $s = 17$. HMN and KH networks have been averaged over $10^2 - 10^3$ independent realizations.

### Random trees

Random trees represent another fundamental class of scale-invariant networks with well-known spectral dimension $d_S = 4/3$. In particular, they are networks that are safely embeddable in space of any dimension $d$. Figure 3(a) shows the correlation dimension for different embedding spaces, while Figure 3(b) shows the scaling of the correlation dimension for an embedding dimension $d = 2$. Note the perfect coincidence with the theoretical expectation. Finally, the 3D representation of a random tree is reported in Figure 3(c).

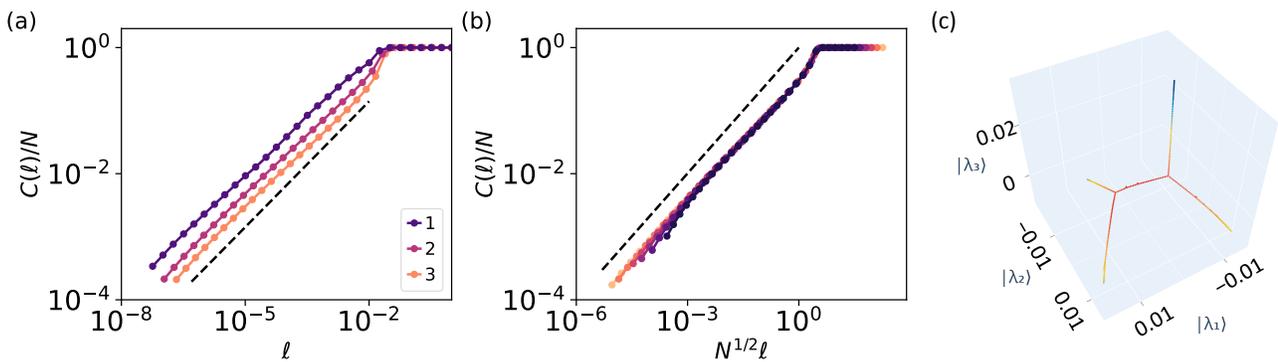

FIG. 3. **Random trees.** **(a)** Correlation integral, $C(\ell)$ versus distance, $\ell$, for different values of $d$ (see legend). Black dashed lines represent the scaling predicted by half of the network spectral dimension, $D = d_s/2 \approx 0.66$. **(b)** Correlation integral, $C(\ell)$ versus normalized distance, $\sqrt{N}\ell$, for a spatial embedding with $d = 2$ and different network sizes, $N = 2^s$ with $s = 10, 11, 12, 13, 14$, and $15$. **(c)** Low-dimensional representation using the three smallest low-frequency normalized network modes, $|\hat{\lambda}_1\rangle$, $|\hat{\lambda}_2\rangle$, and $|\hat{\lambda}_3\rangle$ as the coordinate axes for a random tree of size $N = 2^{13}$.

## 4. ROAD NETWORKS ANALYSIS

Road networks are a specific type of spatial network (i.e., naturally embedded in a 2D Euclidean space) characterized by planar graph properties, where nodes represent intersections and edges represent road segments. Despite their low level of heterogeneity, these networks are another excellent test bed, as their spatial-based heterogeneity provides a controlled environment to evaluate topological regularities in these networks.

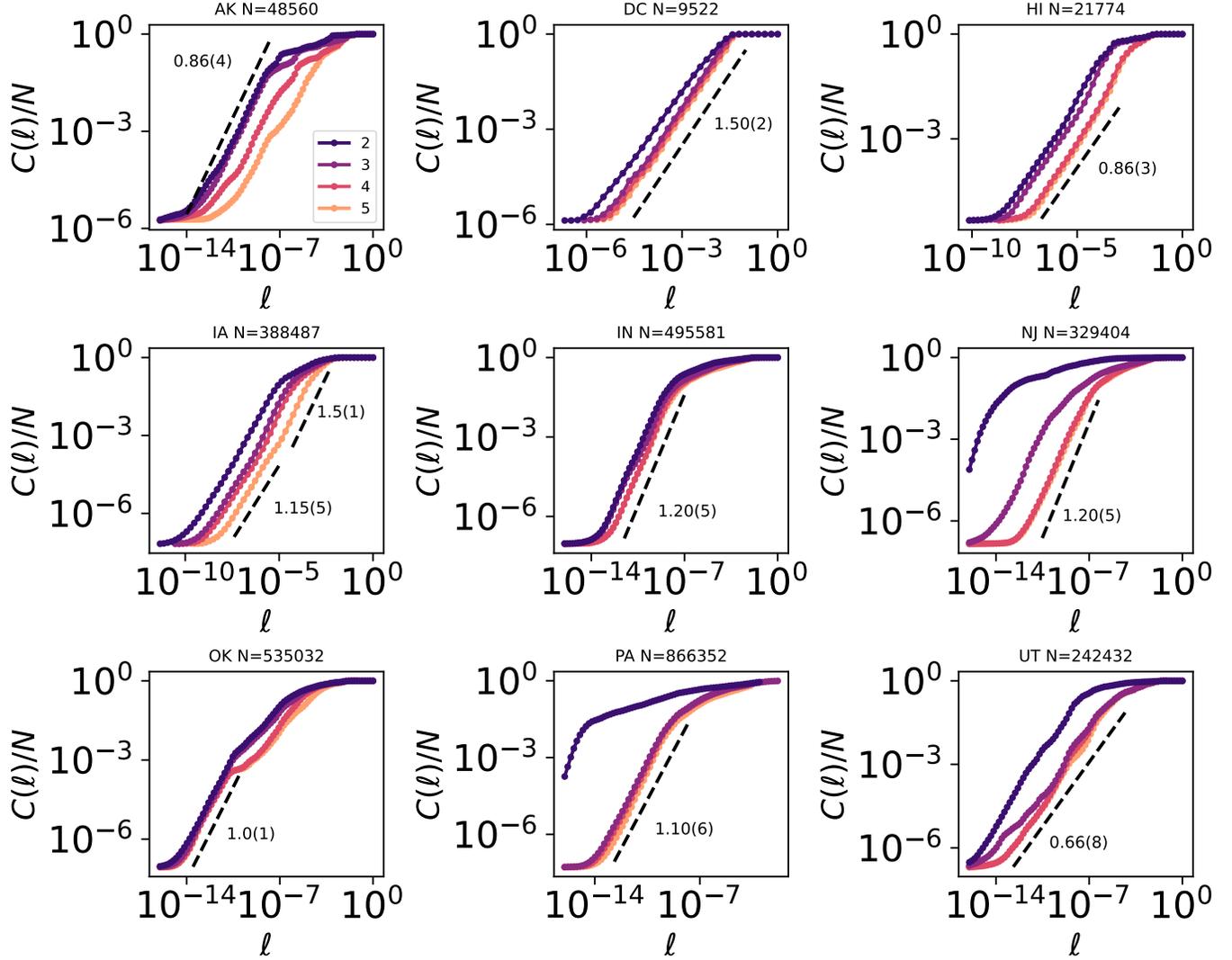

FIG. 4. Correlation integral, $C(\ell)$ versus distance, $\ell$, for different values of $d$ (see legend) and across different U.S. states (see title). The scaling region for all states generally converges for $d > 2$, as expected for this case. Black dashed lines serve as a guide to the eye, representing the fitted network correlation dimension $D$, which is specified for each state.

Figure 4 shows the correlation dimension for road networks across various U.S. states. As observed, for states with less complex topography, such as the District of Columbia (DC), Iowa (IA), or New Jersey (NJ), the correlation dimension is greater than 1. In contrast, for states with more heterogeneous and rugged topography, such as Utah (UT), Hawaii (HI), and Alaska (AK), the correlation dimension yields a value that is expected to be half of the non-integer dimension of the road network, $D = d_s/2$.

Similarly, the European road network, where vertices represent cities and edges represent roads, gives a spatial dimension compatible with $D = 1.5(1)$, as illustrated in Figure 5.

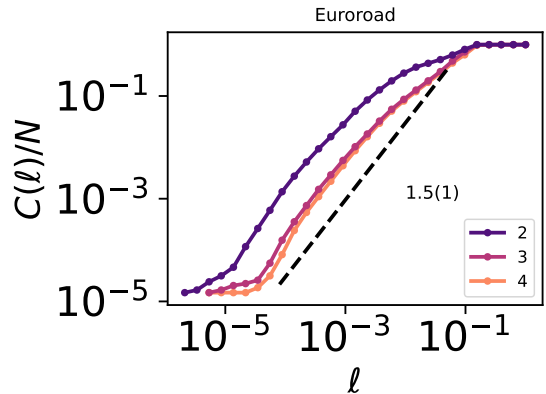

FIG. 5. Correlation integral, $C(\ell)$ versus distance, $\ell$, for different values of $d$ (see legend) for the European Road network. Black dashed lines serve as a guide to the eye, representing the fitted network correlation dimension $D$.

## 5. STOCHASTIC BLOCK MODELS

Analyzing densely populated areas in reciprocal space is crucial for community detection purposes. Figure 6 reports the 2D and 3D embeddings of a Stochastic Block Model with 12 independent modules. Increasing the embedding dimension significantly reduces the overlap between blocks, leading to better separation between the communities. As dimensionality increases, the modules are less likely to overlap, thereby improving the precision in identifying the different modules.

This effect is important because, in lower-dimensional embeddings, communities with similar connection patterns can overlap, making it more difficult to distinguish between them. An important consequence to be explored is the possible connection with transient effects in random walks. As Block Models do not exhibit any finite correlation dimension, the overlap over dimension three can be expected to be negligible, making dimensions $d = 3$ and $d = 4$ ideal for representing clustered networks. However, further work is still needed in this direction.

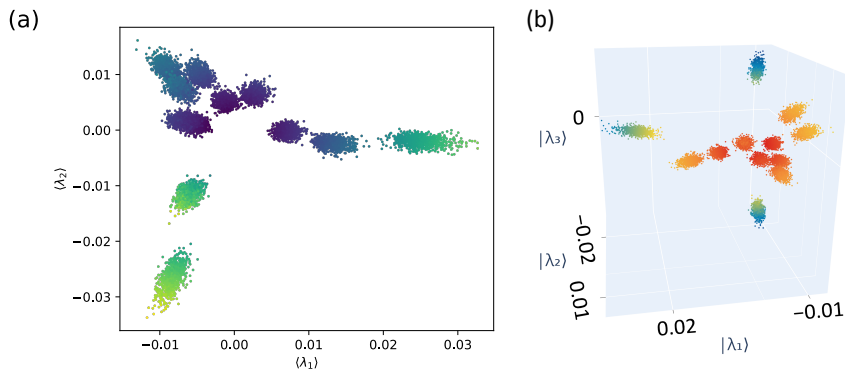

FIG. 6. (a) Low-dimensional representation using the three smallest low-frequency normalized network modes, $|\hat{\lambda}_1\rangle$, $|\hat{\lambda}_2\rangle$, and $|\hat{\lambda}_3\rangle$ as the coordinate axes. (b) Low-dimensional representation using the two smallest low-frequency normalized network modes, $|\hat{\lambda}_1\rangle$, and $|\hat{\lambda}_2\rangle$ as the coordinate axes. Parameters: Stochastic block model with $N = 9996$ nodes and 12 independent modules of size $n = 883$, with intra-module probability $p_1 = 0.25$ and inter-module probability $p_2 = 0.02$.


# REFERENCES

[1] P. Holme and B. J. Kim, Growing scale-free networks with tunable clustering, Phys. Rev. E **65**, 026107 (2002).
[2] P. Moretti and M. A. Muñoz, Griffiths phases and the stretching of criticality in brain networks, Nat. Comm. **4**, 2521 (2013).
[3] S. N. Dorogovtsev, A. V. Goltsev, and J. F. F. Mendes, Pseudofractal scale-free web, Phys. Rev. E **65**, 066122 (2002).
[4] L. Falsi, M. Aversa, F. Di Mei, D. Pierangeli, F. Xin, A. J. Agranat, and E. DelRe, Direct observation of fractal-dimensional percolation in the 3d cluster dynamics of a ferroelectric supercrystal, Phys. Rev. Lett. **126**, 037601 (2021).
[5] F. Xin, L. Falsi, Y. Gelkop, D. Pierangeli, G. Zhang, F. Bo, F. Fusella, A. J. Agranat, and E. DelRe, Evidence of 3d topological-domain dynamics in ktn:li polarization-supercrystal formation, Phys. Rev. Lett. **132**, 066603 (2024).


\end{document}